\documentclass[english,superscriptaddress,twocolumn,aps,prc]{revtex4-1}
\usepackage{graphicx} % Required for inserting images
\usepackage{amsmath}
\usepackage{amsfonts}
\usepackage{comment}
\usepackage{xcolor}
\usepackage{float}
\usepackage{hyperref} 
\hypersetup{
    colorlinks=true,       
    linkcolor=blue,       
    citecolor=blue,        
    urlcolor=blue          
}

\begin{document}

\title{Study of jet-induced hydro response in high-energy heavy-ion collisions with a flow-matching generative model}

\author{Kai-Yi Wu}
\affiliation{Key Laboratory of Quark \& Lepton Physics (MOE) and Institute of Particle Physics, Central China Normal University, Wuhan 430079, China}
\affiliation{Artificial Intelligence and Computational Physics Research Center, Central China Normal University, Wuhan 430079, China}

\author{Zhong Yang}
\email{zhong.yang@vanderbilt.edu}
\affiliation{Department of Physics and Astronomy, Vanderbilt University, Nashville, TN}

\author{Long-Gang Pang}
\email{lgpang@ccnu.edu.cn} 
\affiliation{Key Laboratory of Quark \& Lepton Physics (MOE) and Institute of Particle Physics, Central China Normal University, Wuhan 430079, China}
\affiliation{Artificial Intelligence and Computational Physics Research Center, Central China Normal University, Wuhan 430079, China}

\author{Xin-Nian Wang}
\email{xnwang@ccnu.edu.cn}
\affiliation{Key Laboratory of Quark \& Lepton Physics (MOE) and Institute of Particle Physics, Central China Normal University, Wuhan 430079, China}
\affiliation{Artificial Intelligence and Computational Physics Research Center, Central China Normal University, Wuhan 430079, China}

\begin{abstract}
In high-energy heavy-ion collisions, propagation of the energy deposited into the medium by energetic partons that traverse the quark-gluon plasma (QGP) leads to Mach-cone-like jet-induced medium response. Event-by-event simulations of jet-induced medium responses within  a complete model such as the coupled Linear Boltzmann Transport and hydrodynamic (CoLBT-hydro) model are very resource-intensive. In this study, we develop a flow matching generative model trained by CoLBT-hydro events for the study of the medium response induced by $\gamma$-jets in high-energy heavy-ion collisions. With only the initial spatial and momentum information of the $\gamma$ and jets, the generative model is shown to conditionally reproduce the marginal final-state hadron spectra from the jet-induced hydro response in $0-10\%$ Pb+Pb collisions at $\sqrt{s_{\rm{NN}}}$ = 5.02~TeV. The generative model achieves a computational acceleration  of approximately six orders of magnitude compared to the full CoLBT-hydro simulations, while faithfully preserving the statistical properties of the front and the diffusion wake of the Mach-cone-like hydro response and their contributions to the hadron spectra.  Hadron spectra from the medium response, correlations between the front and diffusion wake and rapidity asymmetry due to the diffusion wake in $\gamma$-hadron correlation are further studied within the generative model.
\end{abstract}


\maketitle

\section{Introduction}

Jets are collimated hadronic flows produced through the fragmentation of high-energy partons (quark or gluon) in high-energy collisions, manifesting highly concentrated clusters of hadrons in the detector. In ultra-relativistic heavy-ion collisions, partons from the initial hard scatterings interact with the quark-gluon plasma (QGP) and lose energy during their propagation in the hot medium, leading to the suppression of high transverse momentum ($p_T$) hadrons and jets,  often referred to as jet quenching~\cite{Baier:2000mf,Gyulassy:2003mc,Kovner:2003zj,Casalderrey-Solana:2007knd,Wiedemann:2009sh,Blaizot:2015lma,Qin:2015srf,Cao:2020wlm,Cao:2024pxc,Wang:2025lct}. The phenomenon can serve as a crucial probe of parton dynamics and properties of the QGP medium in high-energy heavy-ion collisions.
The suppression of  high-$p_T$  hadron and jet cross sections in high-energy heavy-ion collisions has been observed in experiments at RHIC~\cite{PHENIX:2001hpc, STAR:2002ggv, Gyulassy:2004zy, Wang:2004dn} and LHC~\cite{ATLAS:2010isq, ALICE:2010yje, CMS:2011iwn, Majumder:2010qh, Muller:2012zq}, indicating the formation of quark-gluon plasma (QGP) in high-energy heavy-ion collisions. The effective parton and jet energy loss extracted from the experimental data is quite large and depends on the initial jet energy and system size \cite{He:2018gks,Wu:2023azi,Arleo:2022shs}. The energy and momentum lost by jets are deposited into the QGP medium and induce a medium response in the form of Mach-cone-like excitation~\cite{Casalderrey-Solana:2004fdk, Stoecker:2004qu, Ruppert:2005uz, Gubser:2007ga, Qin:2009uh, Li:2010ts} which in turn can modify the internal structure of the jet~\cite{Chen:2021gkj, Yang:2022nei, Yang:2022yfr, Yang:2024ddt, Yang:2024fuv,CMS:2025dua}, affecting many observables, such as jet $R_{AA}$~\cite{He:2022evt, Wang:1998bha, ATLAS:2018gwx, JETSCAPE:2024nkj, JETSCAPE:2022jer, CMS:2021vui, ALICE:2019qyj, Mehtar-Tani:2021fud, Pablos:2019ngg}, jet shape~\cite{JETSCAPE:2025wjn, Tachibana:2017syd, Luo:2018pto, Xiao:2024ffk, Yang:2022nei, CMS:2018jco} and jet fragmentation functions~\cite{Wang:2013cia, Chen:2017zte, Chen:2021gkj, CMS:2021otx, ATLAS:2020wmg, Chien:2015ctp, ATLAS:2018bvp}. 

The interaction involving large-momentum-transfer between hard partons and the QGP medium can be described using perturbative Quantum Chromodynamics (pQCD)~\cite{Guo:2000nz, Zhang:2003wk, Majumder:2009ge}. However, pQCD proves inadequate for describing interaction between soft recoil and thermal medium partons with small momentum transfer which are responsible for the jet-induced medium response. Therefore, jet-induced medium response is typically modeled using the relativistic hydrodynamics with a source term describing the energy and momentum deposition by a propagating jet parton. This approach can accurately capture the dynamical evolution of the medium perturbed by jets and the deposited  energy and momentum. However, it does not consider the effect of the perturbed medium on the subsequent propagation of jet partons. A complete description of jet-medium interactions in high-energy heavy-ion collisions necessitates an integrated framework capable of simultaneously describing both hard and soft degrees of freedom. The coupled Linear Boltzmann Transport and hydrodynamic (CoLBT-hydro) model \cite{Chen:2017zte,Chen:2020tbl,Zhao:2021vmu} based on this idea combines the Linear Boltzmann Transport (LBT) model~\cite{Luo:2023nsi, He:2015pra} and event-by-event (3+1)D CCNU-LBNL viscous hydrodynamic (CLVisc) model~\cite{Pang:2012he, Pang:2012uw, Pang:2018zzo,Wu:2018cpc, Wu:2018pem, Wu:2021fjf} to simultaneously evolve both hard jet and recoil partons and the bulk medium with hydrodynamic response caused by the energy-momentum deposited into the medium by jet partons. This model has successfully described both jet suppression and modification of jet substructures, such as jet shape, fragmentation functions and jet-hadron correlations. The predicted signals of the diffusion wake \cite{Chen:2021gkj,Yang:2022nei,Yang:2025dqu,Yang:2025xni,Yang:2025lii}, a small depletion of particles in the opposite direction of jets, have recently been observed by experiments \cite{CMS:2025dua,ATLAS:2024prm,CMS:2026mur} at LHC. Simulations of these jet events in heavy-ion collisions using the CoLBT-hydro model were carried out in parallelized GPU computers and are very resource-intensive. In order to make the simulations more accessible for more extensive studies, it is essential to develop alternative models that are fast and yet can faithfully describe the coupled evolution of both the jet and medium, including the jet-induced medium response. We develop an AI-based model for this purpose in this study, using data generated from the CoLBT-hydro model.

AI for Science has developed into an active research area, contributing to notable progress across a wide range of scientific disciplines. Among the machine learning methods driving this progress, generative models have attracted particular attention for their ability to learn the underlying statistical distributions of training data and to efficiently sample new data points. Prominent generative frameworks include Variational Autoencoders (VAEs)~\cite{Kingma:2013hel,Doersch:2016eco}, Generative Adversarial Networks (GANs)~\cite{Goodfellow:2014upx}, Transformers~\cite{Vaswani:2017lxt}, Flow-based Models~\cite{Dinh:2014mzt,Dinh:2016pgf}, Flow Matching~\cite{Lipman:2023hef}, and Diffusion Models~\cite{Sohl-Dickstein:2015dhe,Ho:2020epu,Song:2020hus}. Each architecture offers distinct advantages: VAEs are known for their explicit latent space structure. GANs excel at generating high-fidelity samples through adversarial training. Flow-based models, including Flow Matching, provide a mathematically grounded approach to modeling distributions via invertible transformations or vector fields. Unlike GANs, which rely on adversarial training, and diffusion models, which typically generate samples by iteratively reversing a stochastic noise-addition process, Flow Matching learns a continuous vector field that transports samples from a simple prior distribution to the data distribution, often along a straight-line interpolation path. Samples are then generated by solving the corresponding ordinary differential equation, leading to more stable and efficient training while maintaining high sample quality.

In the field of high-energy nuclear physics, machine learning has emerged as a powerful tool to solve many problems~\cite{Pang:2024kid,Ma:2023zfj,Zhou:2023pti,He:2023zin,Boehnlein:2021eym,Zhou:2018ill,Pang:2016vdc,Hirvonen:2023lqy,Stewart:2025vua}. Among these efforts, generative models have been extensively applied across various stages of heavy-ion collisions to accelerate Monte Carlo simulations~\cite{OmanaKuttan:2025qjj}. In Lattice QCD, flow-based models~\cite{Albergo:2019eim,Hackett:2021idh} and diffusion models~\cite{Zhu:2025pmw,Wang:2023exq} serve as efficient samplers for field configurations, which are essential for computing a wide range of QCD observables, including the Equation of State (EoS). Diffusion models have also been combined with complex Langevin dynamics to learn the distributions effectively sampled in theories with a sign problem~\cite{Aarts:2025lpi}. Generative models have successfully accelerated simulations of hydrodynamics~\cite{Sun:2024lgo}. For modeling detector responses, various generative architectures have been developed to perform fast and accurate calorimeter shower simulations~\cite{Krause:2021ilc,Krause:2021wez,ATLAS:2021pzo,Mikuni:2022xry}.

Despite the extensive application of generative models in heavy-ion physics, they have not been applied to accurate simulations of jet-medium interactions, which remain computationally expensive in the traditional approach due to the complexity of coupled evolutions. To address this bottleneck, this study employs the Flow Matching model to accelerate the CoLBT-hydro framework in calculating jet-induced hydro responses. By training on $\gamma$-jet events that incorporate full hydro-response effects, the model learns to effectively predict the final particle spectra from the hydro response directly from the information of the initial $\gamma$-jet configuration, while fluctuations associated with the bulk-medium initial conditions are learned statistically from the training data. To achieve this mapping, a conditional generative approach is essential. Unlike unconditional generation, conditional models synthesize results constrained by specific input variables—in this case, the initial state of the jet. While various architectures support conditional generation, such as the Conditional Invertible Neural Network (CINN) ~\cite{Ardizzone:2019xgg}, Glow Model~\cite{Kingma:2018gyr}, we adopt the Conditional Flow Matching~\cite{Lipman:2023hef} framework. This choice enables us to efficiently bridge the gap between initial parton showers and the final spectra from the hydro response, significantly enhancing computational efficiency while preserving physical fidelity.

In the remainder of this paper, we will first provide a brief introduction to the CoLBT-hydro model and the datasets of $\gamma$-jet events in $0-10\%$ Pb+Pb collisions at $\sqrt{s_{\rm{NN}}}$ = 5.02~TeV from the CoLBT-hydro model simulations. After the description of the Flow Matching model, we will discuss and compare the results from the Flow Matching model and the training data from CoLBT-hydro model. Conclusions will be given at the end.

\section{CoLBT-hydro Model}

The CoLBT model~\cite{Chen:2020tbl,Chen:2017zte} couples the Linear Boltzmann Transport (LBT) model~\cite{He:2015pra,Luo:2023nsi} for energetic partons with the (3+1)D CCNU-LBNL viscous hydrodynamic model (CLVisc)~\cite{Pang:2018zzo,Pang:2012he,Pang:2012uw,Pang:2013pma,Wu:2018cpc,Wu:2018pem,Wu:2021fjf} for the bulk medium evolution. 

\subsection{LBT Model}
In LBT model, the evolution of the phase-space distribution of jet partons, $f_a(p, x, t)$, is governed by the Boltzmann equation:
\begin{align}
    p_a^\mu \partial_\mu f_a = \int \prod_{i=b,c,d} \frac{d^3p_i}{2E_i(2\pi)^3} \frac{\gamma_b}{2}(f_c f_d - f_a f_b) |M_{ab \rightarrow cd}|^2 \notag \\
    \times S_2(\hat{s},\hat{t},\hat{u}) (2\pi)^4 \delta^4(p_a + p_b - p_c - p_d) + \text{inelastic}
\end{align}
where:
\begin{itemize}
    \item $p_a^\mu$ is the four-momentum of parton $a$.
    \item $f_i = (2\pi)^3 \delta^3(\vec{p} - \vec{p}_i) \delta^3(\vec{x} - \vec{x}_i - \vec{v}_i t)$ is the phase-space distribution function of jet-shower partons $(i = a,c)$.
    \item $\gamma_b$ is the color-spin degeneracy (6 for quarks and 16 for gluons). $f_i = 1/(e^{p \cdot u/T} \pm 1)$ is the thermal parton distribution $(i = b,d)$ in the local-equilibrium approximation for quarks ($+$) and gluons ($-$). Here $u^\mu$ is the fluid four-velocity and $T$ is the local temperature. The momenta of thermal partons are sampled in the fluid cell's co-moving frame and then boosted to the laboratory frame.
    \item $f_c f_d$ is the gain term for the $c+d \rightarrow a+b$ process, and $-f_a f_b$ is the loss term for the $a+b \rightarrow c+d$ process. The effect of quantum statistics is neglected in the final state.
    \item $M_{ab \rightarrow cd}$ is the $a+b \rightarrow c+d$ parton scattering amplitude calculated at leading order in perturbative QCD.
    \item $S_2(\hat{s},\hat{t},\hat{u}) = \theta(\hat{s} \ge 2\mu_D^2) \theta(-\hat{s} + \mu_D^2 \le \hat{t} \le -\mu_D^2)$ regularizes collinear divergences, where $\hat{s},\hat{t},\hat{u}$ are Mandelstam variables and $\mu_D$ is the Debye screening mass with $\mu_D^2 = (N_c + N_f/2) g^2 T^2 / 3$. Here $g^2 = 4\pi\alpha_s$ is the strong coupling, $N_c = 3$, and $N_f = 3$ are the numbers of colors and flavors, respectively.
\end{itemize}

The interaction kernels in the Boltzmann equation in the LBT model contain both elastic and inelastic processes between hard partons and medium constituents with perturbative scattering matrices. The total and individual elastic scattering rates for the parton,
\begin{align}
    \Gamma_{\text{el}}^a \equiv \sum_{b,(cd)} \Gamma_{\text{el}}^{ab \to cd},
\end{align}
can be calculated from the perturbative matrices $M_{ab \rightarrow cd}$ \cite{He:2015pra}.

The inelastic scattering rate for induced gluon emission,
\begin{align}
    \Gamma^{a}_{inel} = \int dz dk_{\perp}^2 \frac{1}{1+\delta^{ag}}\frac{dN^a_g}{dz dk_{\perp}^2 dt},
\end{align}
is given by the induced gluon spectrum from the higher-twist formalism~\cite{Guo:2000nz,Zhang:2003wk,Majumder:2009ge},
\begin{align}
    \frac{dN_g^a}{dzdk_{\perp}^2 dt} = \frac{2\alpha_s C_A \hat{q}_a(x) P_a(z) k_{\perp}^4}{\pi \left(k_{\perp}^2 + z^2m^2\right)^4} \sin^2 \left(\frac{t - t_i}{2\tau_f}\right),
\end{align}
where $z$ and $k_\perp$ are the fractional energy and transverse momentum of the emitted gluon, respectively, $\alpha_s$ denotes the strong coupling constant, $m$ is the quark mass, $\hat q_a(x)$ is the jet transport coefficient for parton a, and $C_A$ is the adjoint color factor, $t_i$ is the production time of the jet parton or the last instance of inelastic scattering involving gluon radiation, and $\tau_f$ is the formation time of the radiated gluon, 
\begin{align}
    \tau_f = \frac{2Ez(1-z)}{k_{\perp}^2 + z^2m^2}.
\end{align}
The splitting function $P_a(z)$ for parton $a$ (where $a = q$ for quarks and $a = g$ for gluons) describes the probability of radiating a gluon with energy fraction $z$.
\begin{align}
    P_{q}(z)&=\frac{(1-z)[1+(1-z)^2]}{z}, \\
    P_{g}(z)&=2\frac{(1-z+z^2)^3}{z(1-z)}.
\end{align}

When jet partons interact with the medium during a time interval $\Delta \tau$ in the parton propagation, the scattering probabilities are given by
\begin{align}
    P_{\text{tot}}^a=1-(1-P^a_{\text{el}})(1-P^a_{\text{inel}}) = P_{\text{el}}^a(1 - P_{\text{inel}}^a) + P_{\text{inel}}^a,
    \label{sca_probability}
\end{align}
where
\begin{align}
    P_{\text{el}}^a &= 1 - \exp[-\Delta\tau\Gamma_a^{\text{el}}(x)], \\
    P_{\text{inel}}^a &= 1 - \exp[-\Delta\tau\Gamma_a^{\text{inel}}(x)].
\end{align}
In such a scattering, a thermal parton is sampled from the local fluid cell according to its temperature and the final-state momenta of the scattering are sampled according to the scattering matrices $|M_{ab \rightarrow cd}|^2$. The scattering will also leave a “hole” in the phase space distribution that is treated as a negative parton. After $2\rightarrow n$ scattering, the four-momentum of the jet parton is updated, and the final medium parton is recorded as a recoil parton. These recoil partons, along with radiated gluons, modify the soft-parton distribution around the jet direction and affect the jet's internal structure. More detailed description of LBT can be found in Refs.~\cite{Luo:2023nsi, He:2015pra}.

\subsection{CLVisc Model} 

CLVisc \cite{Pang:2012he, Pang:2012uw, Pang:2018zzo, Wu:2018cpc, Wu:2018pem, Wu:2021fjf} is a (3+1)D viscous hydrodynamic model developed to simulate event-by-event evolution of the QGP formed in high-energy heavy-ion collisions with fluctuating initial conditions. It was written in Open Computing Language (OpenCL) so that it can run in many parallel computing architectures. On GPU parallel computers, it provides a 60-fold speedup for the space-time evolution and a 120-fold speedup for Cooper-Frye particlization relative to that without GPU parallelization. 

The CLVisc model numerically solves the following second-order viscous relativistic hydrodynamic equations,
\begin{align}
     \nabla_{\mu} T^{\mu \nu} =& j^{\nu},
     \label{hydro-eq} \\
    % & \nabla_{\mu} N^{\mu } = 0, \\
     T^{\mu \nu} =& e u^{\mu}u^{\nu} - (P + \Pi)\Delta^{\mu \nu} + \pi^{\mu \nu} 
    % & N^{\mu} = n u^{\mu} + v^{\mu},
\end{align}
where
\begin{itemize}
    \item $j^\nu$ is a source term for the energy-momentum deposition by propagating partons.
    \item $\nabla_{\mu}$ is the covariant derivative, defined as
    \begin{align}
        \nabla_{\mu}A^{\alpha\beta}=\partial_{\mu}A^{\alpha\beta}
+\Gamma^{\alpha}_{\rho\mu}A^{\rho\beta}+\Gamma^{\beta}_{\rho\mu}A^{\alpha\rho}. \notag
    \end{align}
    \item  $e$ is the energy density, $u^{\mu}$ is the flow velocity ($u^{\mu}u_{\mu}=1$), $P$ is the pressure, $\Pi$ is the bulk viscous pressure, and $\pi^{\mu \nu}$ is the shear viscous tensor.
    \item $\pi^{\mu\nu} = \eta \sigma^{\mu\nu} - \tau_\pi [ \Delta^{\mu}_{\alpha} \Delta^{\nu}_{\beta} u^{\lambda} \nabla_{\lambda} \pi^{\alpha\beta} + \frac{4}{3} \pi^{\mu\nu} \theta ] - \lambda_1 \pi_{\lambda}^{\langle\mu} \pi^{\nu\rangle\lambda} - \lambda_2 \pi_{\lambda}^{\langle\mu} \Omega^{\nu\rangle\lambda} - \lambda_3 \Omega_{\lambda}^{\langle\mu} \Omega^{\nu\rangle\lambda}$. \\
    with the expansion rate $\theta=\nabla_{\mu}u^{\mu}$ , symmetric shear tensor $\sigma^{\mu\nu}=2\Delta^{\mu\nu\alpha\beta}\nabla_{\alpha}u_{\beta}$, and the antisymmetric vorticity tensor $\Omega^{\mu\nu}=\frac{1}{2}\Delta^{\mu\alpha}\Delta^{\nu\beta}(\nabla_{\alpha}u_{\beta}-\nabla_{\beta}u_{\alpha})$, tensorial spatial projector: $\Delta^{\mu\nu}_{\alpha\beta} = \frac{1}{2}\left(\Delta^{\mu}_{\alpha}\Delta^{\nu}_{\beta}+\Delta^{\mu}_{\beta}\Delta^{\nu}_{\alpha}\right) - \frac{1}{3}\Delta^{\mu\nu}\Delta_{\alpha\beta}$, and $\tau_{\Pi},\tau_{\pi},\lambda_1,\lambda_2,\lambda_3$ are five independent second-order transport coefficients~\cite{Pang:2018zzo}.
    \item  Projection operator perpendicular to the flow velocity:$\Delta^{\mu \nu}=g^{\mu \nu}-u^{\mu}u^{\nu}$. 
\end{itemize}

In Cartesian coordinates $(t, x, y, z)$, the inverse metric is chosen as $g^{\mu\nu} = \operatorname{diag}(1, -1, -1, -1)$.
Throughout this paper we use the coordinates $X^{\mu} = (\tau, x, y, \eta_s)$, with
\begin{align}
    \tau    &= \sqrt{t^{2} - z^{2}}, \\
    \eta_s  &= \frac{1}{2} \ln \frac{t+z}{t-z}.
\end{align}
From the line element
\begin{align}
    ds^{2} &= dt^{2} - dx^{2} - dy^{2} - dz^{2} \\
           &= d\tau^{2} - dx^{2} - dy^{2} - \tau^{2} d\eta_s^{2} \\
           &= g_{\mu\nu} \, dX^{\mu} dX^{\nu},
\end{align}
together with the transformation $t = \tau \cosh \eta_s$, $z = \tau \sinh \eta_s$, one reads off the covariant metric 
$g_{\mu\nu} = \operatorname{diag}(1, -1, -1, -\tau^{2})$.
Taking the inverse gives the contravariant (inverse) metric
$g^{\mu\nu} = \operatorname{diag}(1, -1, -1, -1/\tau^{2})$, which is the form used in our calculations.

We use the TRENTo-2D \cite{Moreland:2014oya,Bernhard:2016tnd} model to generate the fluctuating initial transverse entropy density profile as the initial conditions for the hydrodynamic equation.  The initial energy density profile is obtained from the entropy density through the equation of state (EoS) assuming local thermal equilibrium.  To account for the longitudinal profile, we apply an envelope function in the spatial rapidity $\eta$,
\begin{align}
H(\eta)=
\begin{cases}
1, & |\eta|<\eta_{\mathrm{flat}}=1.7,\\
\exp\!\left[-\dfrac{(|\eta|-\eta_{\mathrm{flat}})^2}{2\eta_{\mathrm{gw}}^2}\right], & |\eta|\ge \eta_{\mathrm{flat}},
\end{cases}
\end{align}
with $\eta_{\mathrm{gw}}=2.0$. This parametrization gives a plateau around midrapidity and a Gaussian fall-off at large $|\eta|$, providing a longitudinal structure that reproduces the final hadron rapidity distribution.

When solving the above hydrodynamic equations, the default s95p-pce EoS is used. It assumes partial chemical equilibrium given by the lattice
QCD EoS \cite{Huovinen:2009yb} at high energy density and hadronic resonance gas (HRG) EoS at low energy density with a smooth crossover in between using interpolation. Zero bulk viscosity and a constant shear viscosity to entropy ratio $\eta/s=0.15$ are used which is constrained through comparisons to experimental data on hadron spectra in heavy-ion collisions \cite{Pang:2018zzo}. The final hadrons are sampled according to the Cooper-Frye freeze-out on the hypersurface at a temperature $T_f=137$ MeV.

\subsection{CoLBT Model}

In the CoLBT-hydro model, the coupling between LBT and CLVisc model is achieved by introducing a source term $j^\nu$ in the hydrodynamic equations Eq.~(\ref{hydro-eq}),
\begin{align}
    j^\nu =& \sum_i \frac{\theta(p_{\text{cut}}^0 - p_i \cdot u) p_i^\nu / \Delta\tau}{\tau(2\pi)^{3/2} \sigma_r^2 \sigma_{\eta_s}}\notag \\ \times &\exp\left[ -\frac{(\vec{x}_\perp - \vec{x}_{\perp i})^2}{2\sigma_r^2} - \frac{(\eta_s - \eta_{si})^2}{2\sigma_{\eta_s}^2} \right],
\end{align}
which is determined by the four-momentum flow from soft partons whose energy in the local co-moving frame is below the cut-off $p^0_{\rm cut}=2$ GeV and all the negative partons. A Gaussian smearing is used for each soft and negative parton at the transverse $\vec{x}_{\perp,i}$ and spatial rapidity $\eta_{si}$ coordinates, with the width $\sigma_r$ and $\sigma_{\eta_s}$. In practice, this amounts to taking all the negative partons and soft partons below $p^0_{\rm cut}$ out of LBT transport at each time step and using them to construct the source term. The hydrodynamic equations with this source term are evolved for this time step and provide the space-time profile of the local temperature and fluid velocity which will be used for the LBT transport for the next time step. 

This real-time coupling allows the concurrent hydrodynamic evolution of the bulk medium and transport of jet shower and hard recoil partons, leading to the formation of Mach cones and diffusion wakes~\cite{Yang:2022yfr}. When the local temperature drops below the QCD phase transition temperature, $T_{\mathrm{c}} = 165~\mathrm{MeV}$, the interaction between hard partons and the medium stops while the bulk medium continues to evolve according to CLVisc without a source term until the freeze-out temperature, $T_{\mathrm{f}} = 137~\mathrm{MeV}$, when final hadrons are generated from the freeze-out hypersurface according to a Monte Carlo sampling approach, which provides the four-momentum of the final hadrons. The hard partons will hadronize according to a fragmentation-recombination model \cite{Zhao:2021vmu,Zhao:2020wcd}.

\section{Training Datasets}

\label{sec:data}
To simulate $\gamma$-jet events for the training data, we use  PYTHIA 8 \cite{Sjostrand:2014zea} to generate the initial photon and jet parton shower configurations. We also use FASTJET anti-$k_T$ algorithm \cite{Cacciari:2008gp} to reconstruct jets from these jet partons. We use the kinematics of the first three jets with $p_T^\text{jet}>10$~GeV  
and the momentum of the photon with $p_T^\gamma>100$~GeV as a global tag of the initial $\gamma$-jet events. The jet pseudorapidity cut is $|\eta_{jet}| \le 1.6$.

The initial production positions of the $\gamma$-jets are sampled according to the binary collision distribution in the TRENTo-2D model for the initial conditions of the hydrodynamic evolution. These jet shower partons together with the TRENTo-2D initial conditions are the starting point for the CoLBT-hydro model simulations.

The particle spectra of jet-induced hydro response are obtained by subtracting the hydrodynamic background, which is from CLVisc simulations with the same initial conditions for the bulk matter but without jets, from the CoLBT-hydro simulations with jets. To simplify the machine learning process, we rotate all data so that the photon's azimuthal angle is set to $\phi = 0$, and the jet's azimuthal angle is around $\phi = \pi$. 

We generated the training dataset for 16,000 $\gamma$-jet events from CoLBT-hydro simulations in the $0-10\%$ centrality class of Pb+Pb collisions at $\sqrt{s_{\rm{NN}}}$ = 5.02~TeV with $\gamma$-jet initial configurations from Pythia 8 and 400 hydro initial conditions according to the TRENTo-2D model cycled among the 16000 events. The final particle spectra $d^3N/dp_Td\eta d\phi$ in $p_T \in [0,4]$~GeV/c, $\eta \in [-2.7,2.7]$, $\phi \in [0,2\pi]$ for these 16,000 events will be used as the training dataset.

The training dataset is tagged by the information of the photon and jets with 14 input variables: the photon's three-momentum $(p_T^\gamma,\eta^\gamma,\phi^\gamma)$, the transverse position of the initial $\gamma$-jet production $(x,y)$ and the three-momentum  $(p_T,\eta,\phi)$ for each of the three leading jets, with zero-padding applied where necessary.  For events with fewer than three jets with $p_T^\text{jet}>10$~GeV, missing jet quantities were set to zero. These initial partonic jets are reconstructed from final-state partons from Pythia 8 using anti-$k_T$ FASTJET algorithm \cite{Cacciari:2011ma}. Note that we do not tag the initial entropy density distribution for the hydrodynamic evolution. Therefore, event-by-event fluctuations caused by the initial conditions for the hydro evolution are implicitly embedded in the training data.

Principal Component Analysis (PCA)~\cite{doi:10.1098/rsta.2015.0202} was applied to compress each training sample of the 3D spectrum $d^3N/dp_T\,d\eta\,d\phi$ into a lower-dimensional feature vector. Specifically, each spectrum was first flattened into a vector $x_n \in \mathbb{R}^D$, where $n=1,\dots,M$ labels the training samples, $M$ is the total number of samples, and $D$ is the number of bins in the discretized 3D spectrum. Collecting all samples row by row gives the data matrix
\begin{align}
X =
\begin{pmatrix}
x_1^T\\
x_2^T\\
\vdots\\
x_M^T
\end{pmatrix}
\in \mathbb{R}^{M\times D},
\end{align}
where $x_n \in \mathbb{R}^D$. The sample mean of the dataset is computed as
\begin{align}
\mu = \frac{1}{M}\sum_{n=1}^{M} x_n \in \mathbb{R}^D,
\end{align}
and each sample is mean-centered according to
\begin{align}
\tilde x_n = x_n - \mu,
\end{align}
or equivalently, in matrix form,
\begin{align}
\tilde X = X - \mathbf{1}\mu^T,
\end{align}
where $\mathbf{1}\in\mathbb{R}^{M}$ is a column vector with all entries equal to 1.

The covariance matrix of the centered data is then constructed as
\begin{align}
C = \frac{1}{M-1}\tilde X^T \tilde X \in \mathbb{R}^{D\times D}.
\end{align}
The principal directions are obtained by solving the eigenvalue problem
\begin{align}
C w_i = \lambda_i w_i,
\end{align}
where the eigenvectors $w_i$ are ordered according to descending eigenvalues,
\begin{align}
\lambda_1 \ge \lambda_2 \ge \cdots \ge \lambda_D.
\end{align}
We compress the spectrum dataset by keeping only the first $K=N_{\rm medium}=50$ eigenvectors and form the projection matrix
\begin{align}
W = \left(w_1, w_2, \dots, w_K\right)\in\mathbb{R}^{D\times K}.
\label{latent_matrix}
\end{align}
The corresponding low-dimensional PCA representation is then
\begin{align}
z_n = W^T(x_n-\mu),
\end{align}
or, for the whole dataset,
\begin{align}
Z = (X-\mathbf{1}\mu^T)W.
\end{align}
These $K$-dimensional PCA coefficients were normalized to the range $[-1,1]$ and used as the input/output representation of the network. After the network predicts the reduced feature vector, the original 3D spectrum is reconstructed by first undoing the normalization and then applying the inverse PCA transformation,
\begin{align}
\hat x_n = W z_n + \mu,
\end{align}
or, in matrix form,
\begin{align}
\hat X = ZW^T + \mathbf{1}\mu^T.
\end{align}
Here, $\hat X$ denotes the reconstructed approximation to the original data $X$.
\section{Flow Matching model}

\begin{figure*}[ht]
    \centering
    \includegraphics[width=0.8\linewidth]{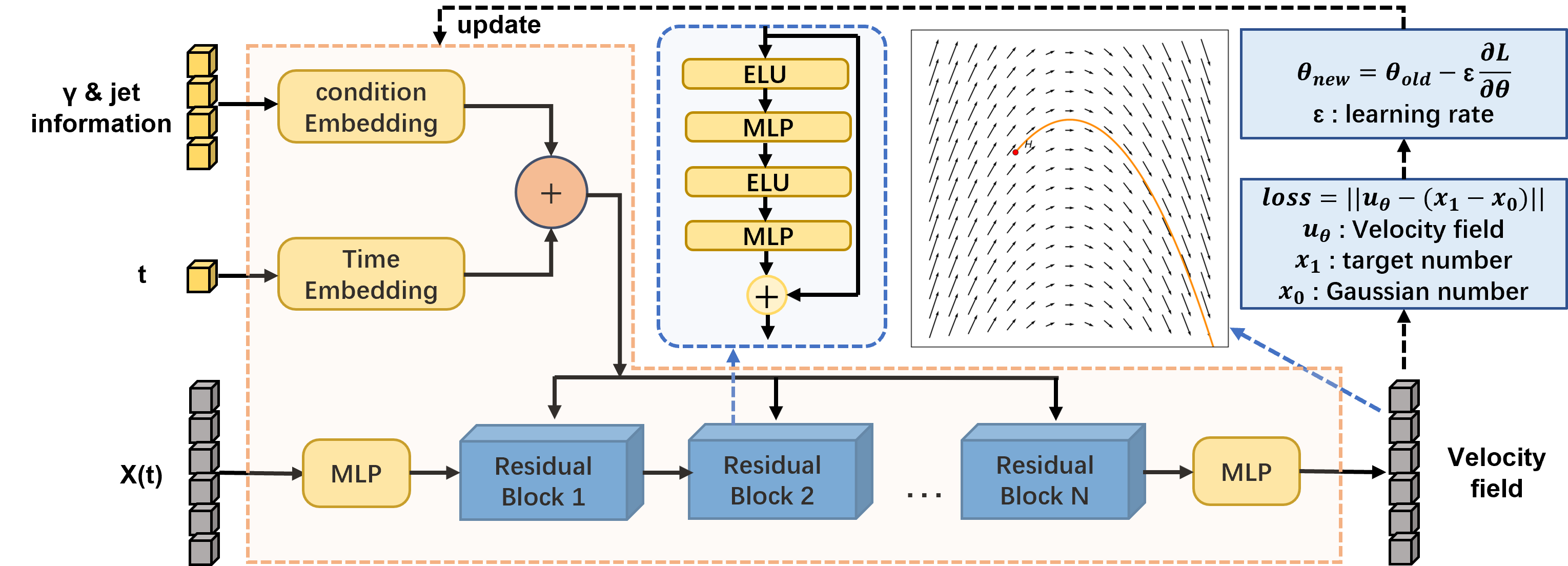}
    \caption{Conditional Flow Matching network structure}
    \label{fig:flowstructure}
\end{figure*}

Generative models learn the distribution of a dataset in order to generate new samples from it. Normalizing flow~\cite{Dinh:2014mzt,Dinh:2016pgf,Gao:2020vdv} and flow matching models~\cite{Lipman:2023hef} are analogous to the transformation sampling method in computational physics: one samples from a simple distribution and applies a set of transformations to obtain samples from a complex target distribution that is otherwise difficult to sample. Flow matching achieves this by learning a vector field that acts as a velocity, guiding sample trajectories in a high-dimensional space. Each sample moves like a particle along a smooth trajectory governed by an ordinary differential equation (ODE), effectively transporting it from a simple source distribution to the target distribution. Once trained, generating a new sample reduces to drawing from a simple source (e.g., a high-dimensional Gaussian) and integrating the learned ODE.

We built a Conditional Flow Matching model that generates event-by-event medium response conditioned on the initial photon ($\gamma$) and jet information. This is non-trivial because the generated samples must not only obey the imposed condition but also carry diverse physical fluctuations: the stochastic number of parton–medium collisions while traversing the QGP, variations in the collision vertices along the jet trajectory, and fluctuations in the energy–momentum transfer during individual scatterings. The model further accounts for inhomogeneities in the QGP energy density background and the intrinsic randomness of Monte Carlo hadronization (Cooper–Frye sampling). All these fluctuation sources are encoded in a random vector drawn from a high-dimensional Gaussian distribution. After training, the fluctuations inherited from this simple distribution should replicate the complex physical fluctuations of our Monte Carlo model.

Fig.~\ref{fig:flowstructure} illustrates the architecture of our Conditional Flow Matching framework. 
The governing Ordinary Differential Equation (ODE) is
\begin{align}
    \frac{dx}{dt}=u_{\theta}(x,t,C),
\end{align}
where $x_0 \equiv x(t=0)$ represents samples from a simple source distribution (50-dimensional Gaussian), and $x_1 \equiv x(t=1)$ represents samples from the target distribution (the first 50 principal components obtained via PCA). Both are accessible during training. 
The time parameter $t \in [0, 1]$ controls the flow from the initial to the final distribution. Here $C$ denotes the condition vector that encodes the initial $\gamma$-jet configuration. The velocity field $u_{\theta}(x,t,C)$ is parameterized by a deep residual neural network with trainable parameters $\theta$, and its output has the same dimensionality as $x_0$ and $x_1$.

We adopt the widely used linear interpolation between source and target,
\begin{align}
    x(t) = t x_1 + (1-t) x_0,
\end{align}
whose time derivative is simply 
\begin{align}
    \frac{dx}{dt} = x_1 - x_0.
\end{align}
This directly provides the training objective, minimizing the loss
\begin{align}
    \mathcal{L}(\theta)=\left\lVert u_{\theta}(x,t,C)-(x_1-x_0)\right\rVert.
\end{align}

The input to $u_{\theta}(x,t,C)$ comprises three components: the linearly interpolated $x(t)$, the time $t$, and the conditions (specifically $\gamma$ and jet information). 
In practice, the 50-dimensional $x(t)$ is first fed into an MLP (multilayer perceptron, a simple feed forward neural network) to produce a 64-dimensional latent vector. 
Simultaneously, $t$, $\gamma$, and jet information are passed through two distinct neural networks, yielding two 64-dimensional embeddings. These three vectors are then added element-wise, forming a 64-dimensional input to a residual block~\cite{He:2015wrn}. 

The internal structure of each residual block is shown in the blue dashed box in Fig.~\ref{fig:flowstructure}. 
The nonlinear activation function used in the block is the exponential linear unit (ELU)~\cite{Clevert:2015qvd}, defined as
\begin{align}
\mathrm{ELU}(x)=
\begin{cases}
x, & x>0,\\
\alpha \left(e^x-1\right), & x\le 0.
\end{cases}
\end{align}
The input-output relation of a residual block is
\begin{align}
h_{l+1} = h_l + f(h_l;\theta_l),
\end{align}
where $h_l$ and $h_{l+1}$ denote the input and output of the $l$-th residual block, respectively, and $f(h_l;\theta_l)$ represents the nonlinear transformation parameterized by $\theta_l$.

During backpropagation, the gradient with respect to the block input satisfies
\begin{align}
\frac{\partial \mathcal{L}}{\partial h_l}
=
\frac{\partial \mathcal{L}}{\partial h_{l+1}}
\frac{\partial h_{l+1}}{\partial h_l}
=
\frac{\partial \mathcal{L}}{\partial h_{l+1}}
\left(
I + \frac{\partial f(h_l;\theta_l)}{\partial h_l}
\right),
\end{align}
where $\mathcal{L}$ is the loss function and $I$ is the identity matrix. The identity shortcut therefore provides a direct path for gradient propagation, which helps mitigate the vanishing-gradient problem and makes deep networks easier to optimize. 
The residual block is repeated $N=6$ times with the same design, and the final latent vector is decoded by another MLP into a 50-dimensional output.

For each pair $(x_0, x_1)$, the network outputs $u_{\theta}(x,t,C)$ and the loss $\mathcal{L}$ is evaluated. 
The parameters $\theta$ are optimized using Adam~\cite{Kingma:2014vow} (learning rate $0.001$), a first-order adaptive stochastic optimization method, with batch size 32. 
Once trained, new samples are obtained by evolving an initial $x_0$ to $x_1$ via the ODE, solved with a second-order Runge-Kutta integrator.

\subsection{Training details and inference procedure}

Given the model construction above, we summarize here the practical
training setup and the reconstruction procedure used in the generation stage.
A total of 16000 events are used in this study, among which 12000 events are
used for training and the remaining 4000 events are used for validation. The model is trained for
5000 epochs. To improve numerical stability, gradient clipping with a maximum
norm of 5.0 is applied after backpropagation. The learning rate is adjusted
according to the validation loss using the \texttt{ReduceLROnPlateau}
scheduler with factor 0.5, patience 50, and minimum learning rate $10^{-10}$.
During training, we monitor both the training and validation losses and retain
the checkpoint whenever they simultaneously improve with respect to the
previous best record. The retained checkpoint is then used in the subsequent
generation procedure.

For inference, the latent sample defined above is evolved from $t=0$ to $t=1$
using 1000 uniform time steps with the same second-order Runge--Kutta solver
introduced above. The generated vector at the final time remains in the
normalized PCA space. Therefore, after the time evolution, each generated
component is first transformed back to the original PCA scale through
\begin{equation}
x_{\mathrm{pca}}
=
\frac{x_{\mathrm{norm}}+1}{2}\odot
\left(x_{\max}-x_{\min}\right)
+x_{\min},
\end{equation}
where $x_{\min}$ and $x_{\max}$ denote the component-wise minimum and maximum
values used in the normalization step. The full spectrum is then reconstructed
through the inverse PCA transformation,
\begin{equation}
X_{\mathrm{recon}} = x_{\mathrm{pca}} W^T + \mu,
\end{equation}
where $W$ is the PCA projection matrix defined in Eq.~(\ref{latent_matrix}), and $\mu$ is the mean vector, respectively. Finally, $X_{\mathrm{recon}}$ is reshaped back to the original
$(\eta,p_T,\phi)$ grid to obtain the generated hydro-response spectrum.

\section{Model Comparisons}
To validate the Flow Matching model, we compare the results of the hydro response events generated by the Flow Matching model with those from the CoLBT-hydro training data with the same initial $\gamma$-jet configurations (4-momentum and production transverse position). 

\subsection{Front and Diffusion wake as hot and dark spots}

Before comparing event-averaged hadron spectra and their fluctuations, we first present a visual comparison of randomly sampled individual events from the training dataset against the corresponding events generated from the Flow Matching model. For each given initial condition of the $\gamma$-jet configuration, a single generation involves sampling one data point from a high-dimensional distribution space (in this case, 50 dimensions).

\begin{figure*}[t] 
  \centering
  \begin{minipage}[b]{0.48\linewidth} 
    \centering
    \includegraphics[width=\linewidth]{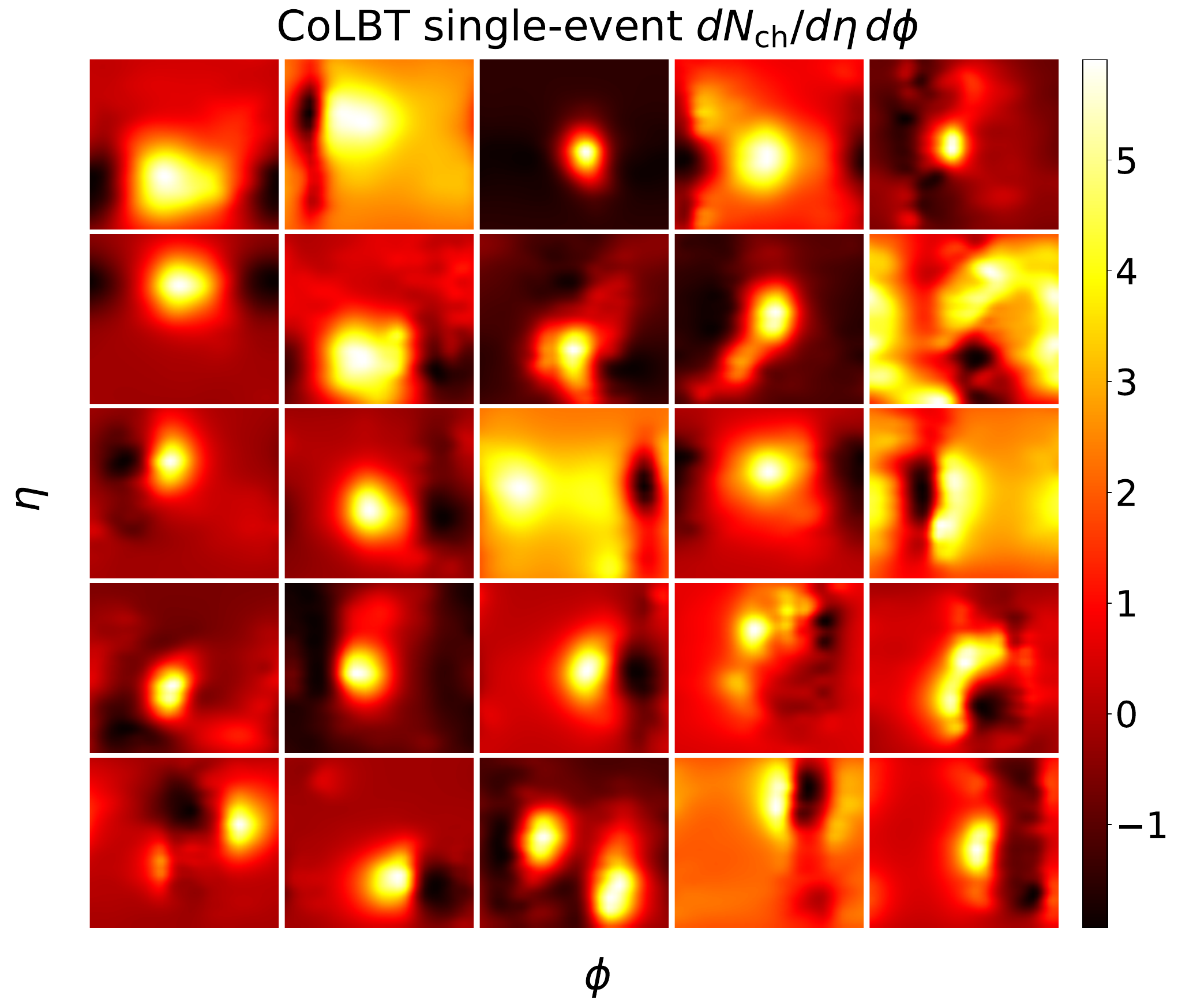} 
  \end{minipage}
  \hfill 
  \begin{minipage}[b]{0.48\linewidth}
    \centering
    \includegraphics[width=\linewidth]{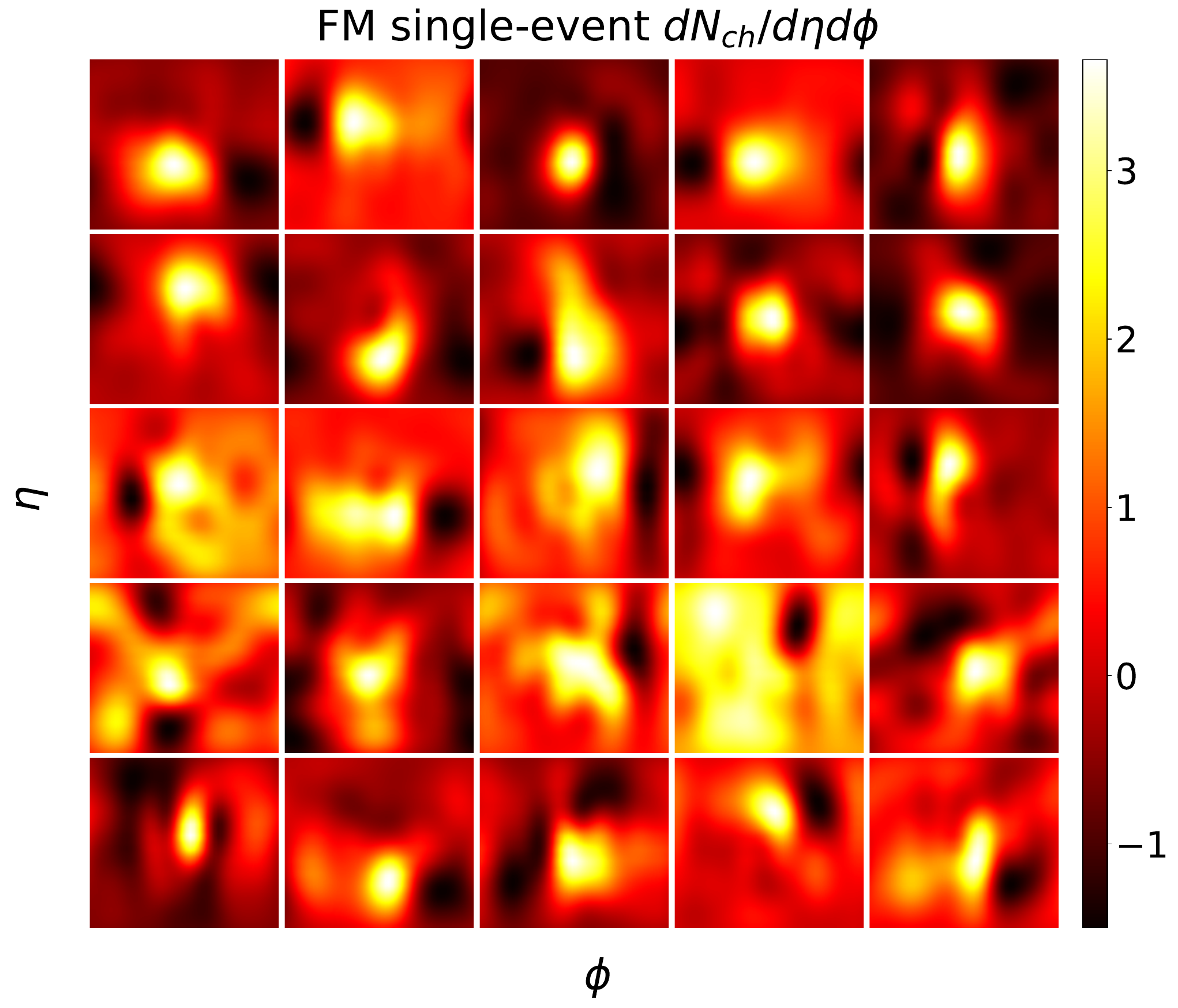} 
  \end{minipage}
  \begin{minipage}[b]{0.48\linewidth}
    \centering
    \includegraphics[width=\linewidth]{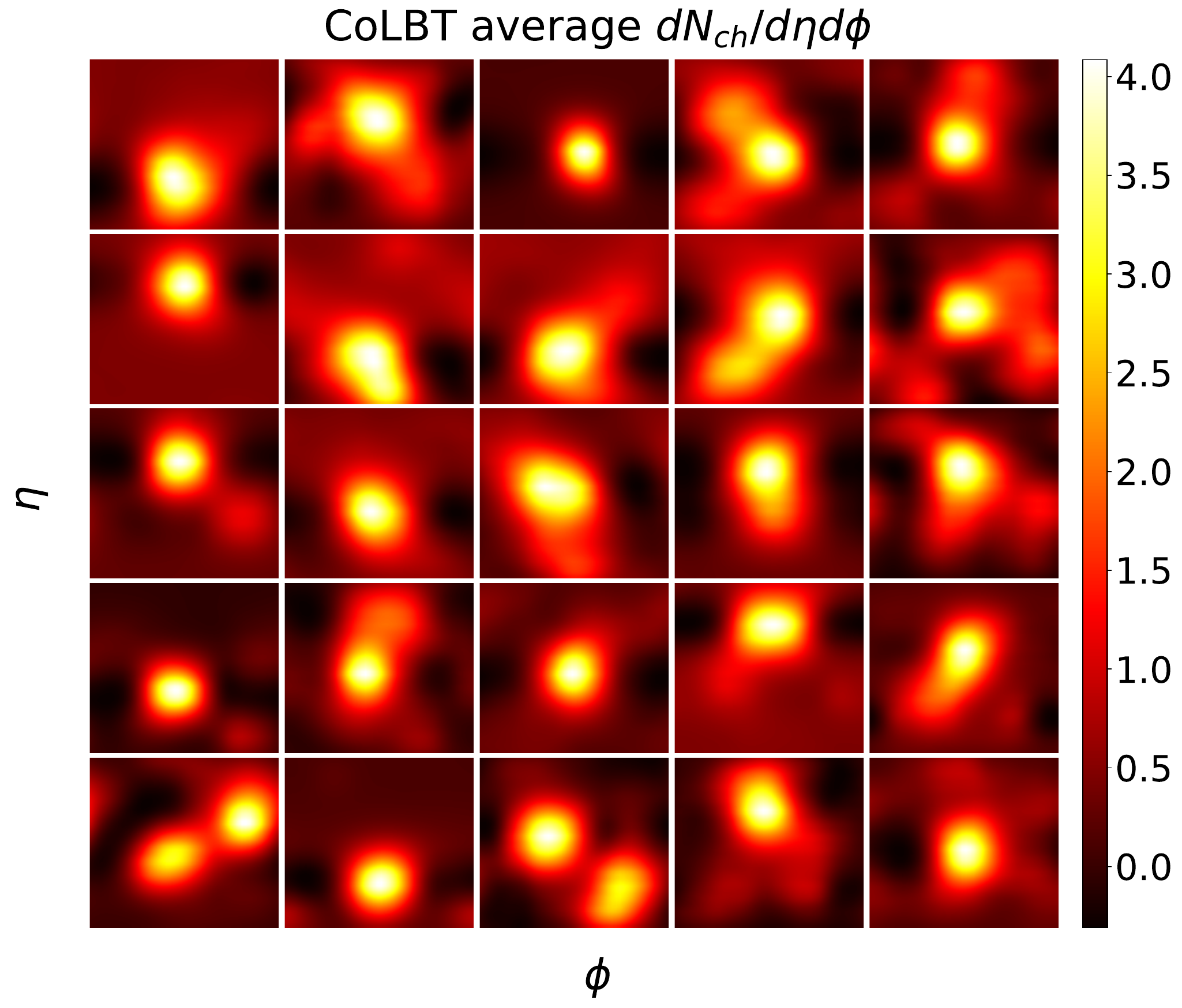}
  \end{minipage}
  \hfill
  \begin{minipage}[b]{0.48\linewidth}
    \centering
    \includegraphics[width=\linewidth]{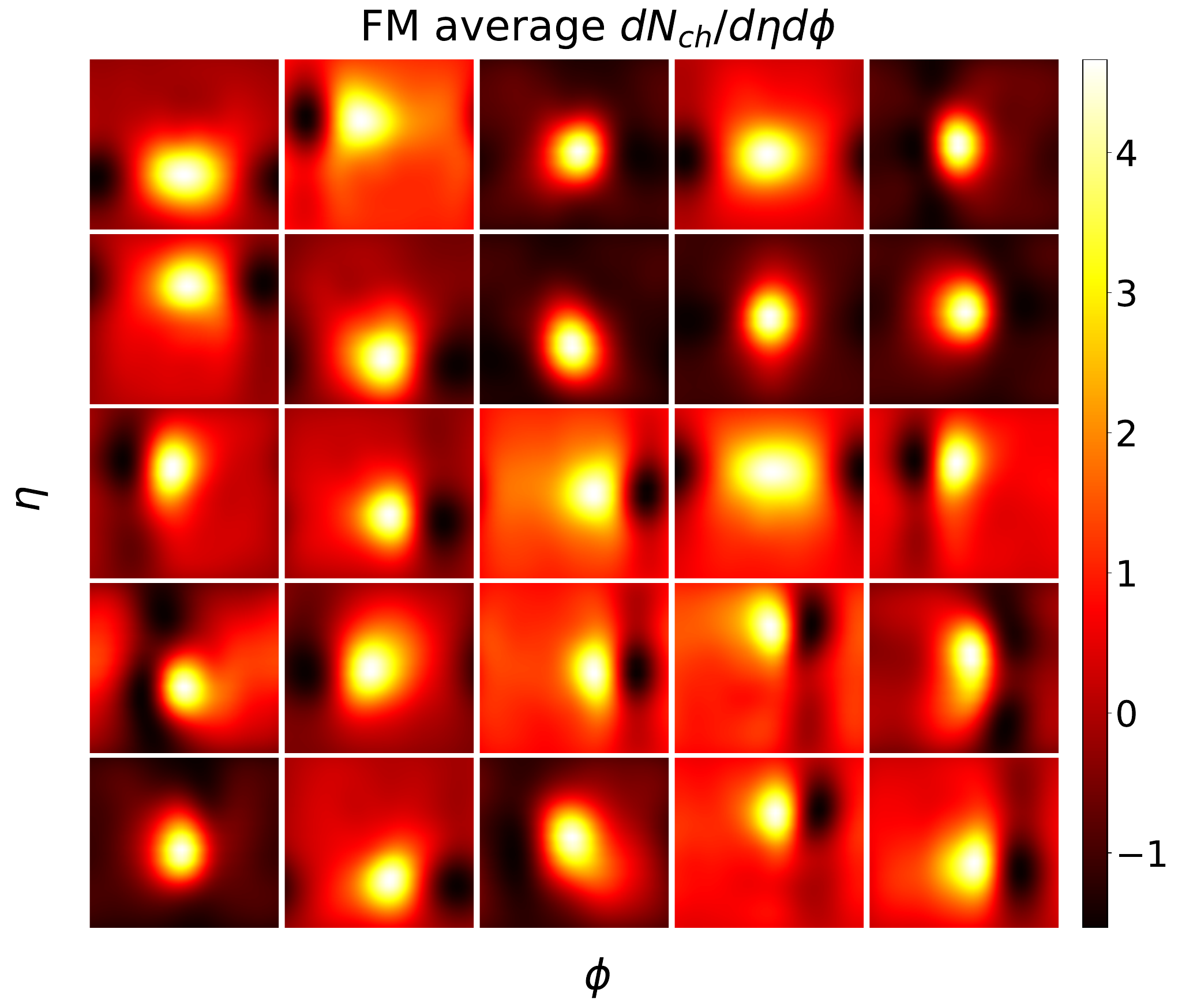}
  \end{minipage}

  \caption{Upper-Left: Single-event $d^2N/d\eta d\phi$ of charged hadrons from the hydro response induced by $\gamma$-jets from CoLBT-hydro simulations of $0-10\%$ central Pb+Pb collisions at  $\sqrt{s_{\rm{NN}}}$ = 5.02~TeV with 25 randomly sampled initial conditions. Upper-Right:  The same as Upper-Left panel except from Flow Matching model generations. Lower-Left: The same as the Upper-Left panel except averaged over 100 simulations for each initial condition. Lower-right: The same as Lower-Left panel except from Flow Matching model generations. The vertical axis is $\eta$, ranging from $[-2.76,2.76]$. The horizontal axis is $\phi$, ranging from $[0,2\pi]$. The spectra are integrated over the transverse momentum range $p_T \in [0,4]$ GeV/$c$.}
  \label{3Dmultishow0}
\end{figure*}

In the upper-left panel of Fig.~\ref{3Dmultishow0}  we show charged hadron distributions $dN_{\rm ch}/d\eta d\phi$ from 25 events randomly selected from the initial training dataset. In each subplot, the vertical axis represents pseudo-rapidity ($\eta \in [-2.76,2.76]$), and the horizontal axis represents the azimuthal angle ($\phi \in [0,2\pi]$). Note again that these distributions are generated by subtracting a jet-free background spectrum (hydro events with the same initial condition but without $\gamma$-jet) from the spectrum in events with jets. Shown in the upper-right panel are  charged hadron distributions from 25 corresponding events generated by the Flow Matching model with the same initial conditions for $\gamma$-jet configurations. 

In each panel, hot (bright) spots correspond to the front wake of the Mach cone structure induced by jets, where the energy deposited into the medium by the propagating jet leads to an enhanced particle number density over the hydro-background without the jet. Conversely, the dark spots (regions) represent the diffusion wake, regions of negative particle density (depletion of particle number density as compared to the hydro events without jets) that forms behind the front wake due to the back-reaction in the Boltzmann transport. While the front and diffusion wakes appear in the same region of the pseudo-rapidity $\eta$, they are oriented back-to-back in the azimuthal  angle ($\phi$). In principle, this would result in an angular separation of $\Delta\phi = \pi$ between the hot and dark spots. However, as one can observe, the actual angular separation often deviates from $\pi$. This is attributed to various effects, such as the local velocity of the fluid and gradient of the medium. We have rotated each event so that $\gamma$ is in the $\phi=0$ direction. Hot spots due to the front wake in the jet direction in these plots are concentrated near $\phi = \pi$. Since jets are distributed symmetrically around $\eta=0$, the front (hot spots) and diffusion wakes (dark spots) also follow the same distribution pattern in $\eta$.

Since event-by-event fluctuations can arise from initial bulk matter distribution, the jet propagation within the CoLBT model as well as in stochastic conditional generation in the Flow Matching model, one should not expect precise one-to-one correspondence between the CoLBT-hydro event (upper-left panel) and the Flow Matching model event (upper-right panel) even though both have the same initial conditions for $\gamma$-jet configuration. The difference reflects the fluctuations in both the training data (CoLBT-hydro model) and the Flow Matching model generation. The correspondence can be approximately established only if one averages over many events for each initial $\gamma$-jet configuration as shown in the lower-left and lower-right panel where the charged hadron distributions for the same 25 initial $\gamma$-jet configurations are averaged over 100 events in both CoLBT-hydro simulations and the Flow Matching model generations.
Since we do not tag the initial conditions for the bulk hydro, the Flow Matching model results are essentially averaged over the hydro initial conditions. For comparison, the event-average of the CoLBT-hydro results on the lower-left panel is also over 100 different hydro initial conditions for each $\gamma$-jet initial configuration. We also compared to the CoLBT-hydro model results where the average is over 100 events with identical initial hydro conditions and found the difference due to the fluctuation of the initial hydro transverse profile small.

Visual inspection of these generated 2D spectra confirms that the model is capable of discerning the averaged positions and sizes of the front and diffusion wake for the same given initial $\gamma$-jet configuration. Nevertheless, some discrepancies are observed in a small subset of events, indicating certain limitations in the current model. For example, rare two-sub-jet events in the training dataset are not learned by the Flow Matching model. These point to the necessity of exploring improved network architectures in the future to enhance the model's ability to cover rare events.

\begin{figure}[h]
    \centering
    \includegraphics[width=0.8\linewidth]{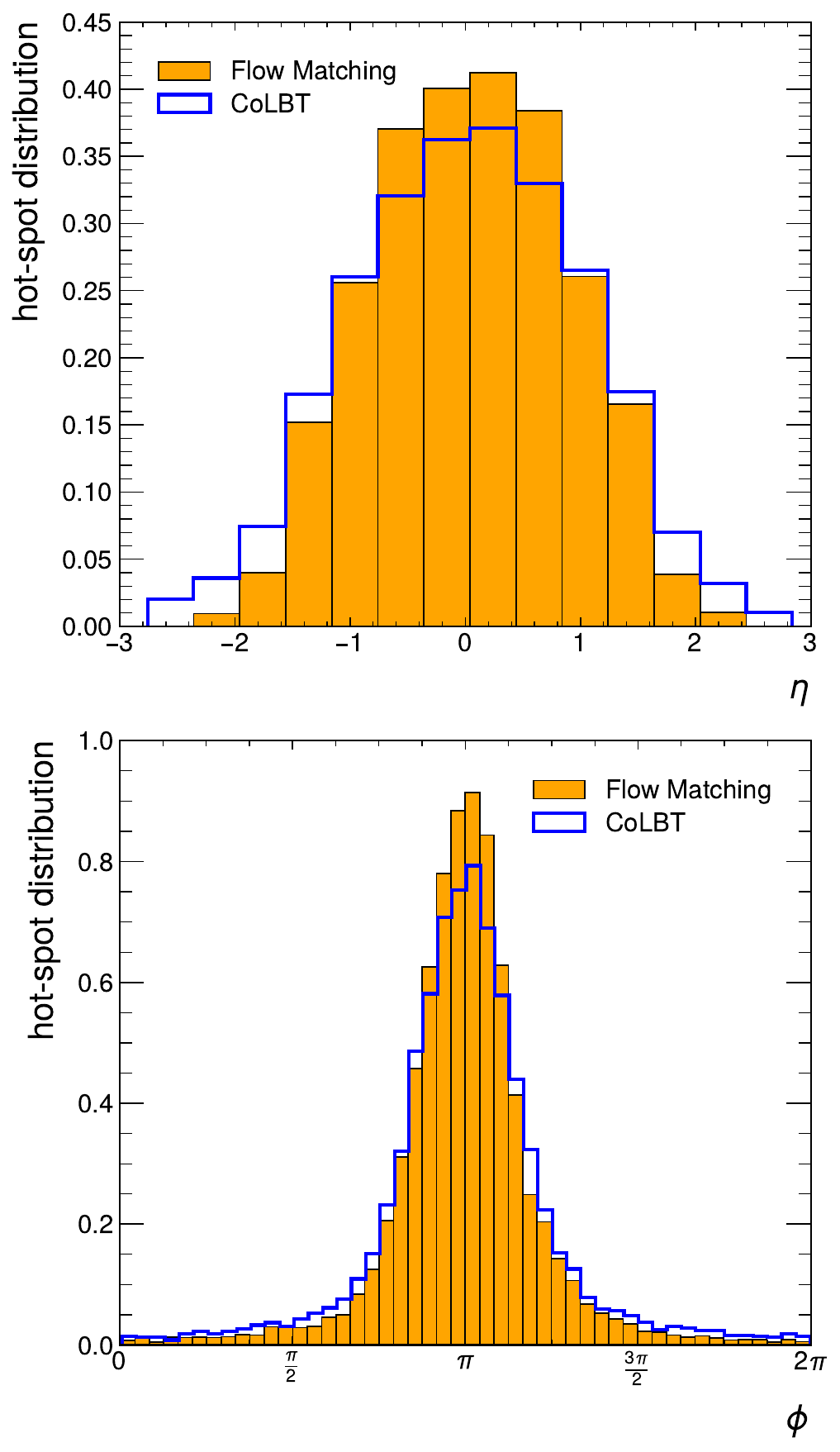}
    \caption{The location distributions of hot spots of front wakes in rapidity $\eta$ (upper) and azimuthal angle $\phi$ (lower) from the Flow Matching model (shaded histograms) as compared to the CoLBT-hydro (open histograms) simulations.}
    \label{3Dfrontwake}
\end{figure}

To quantify the comparisons between the 2D hadron spectra from the hydro response to jet propagation from CoLBT-hydro and Flow Matching model we plot in Fig.~\ref{3Dfrontwake} the distributions of the locations of the front wake hot spots in rapidity and azimuthal angle. The locations are defined as the points of maximum hadron number density from the hydro response in each event. One can see that the Flow Matching model can quantitatively learn the structure of the 2D hadron spectrum and its fluctuations. The location distributions of the hot spots, which are centered at $\eta=0$ and $\phi=\pi$ (opposite direction to the photon), are expected to align with the jets, whose directions deviate from $\eta=0$ and $\phi=0$ due to the $\gamma$-jet kinematics and initial state radiation, leading to the broad distributions seen in Fig.~\ref{3Dfrontwake}. The distributions from the Flow Matching model (shaded histogram) are slightly narrower than the training data (open histogram) in both pseudo-rapidity and azimuthal angle due to its inability to learn the structure of rare events with two or more sub-jets. 

\begin{figure}[h]
    \centering
    \includegraphics[width=0.8\linewidth]{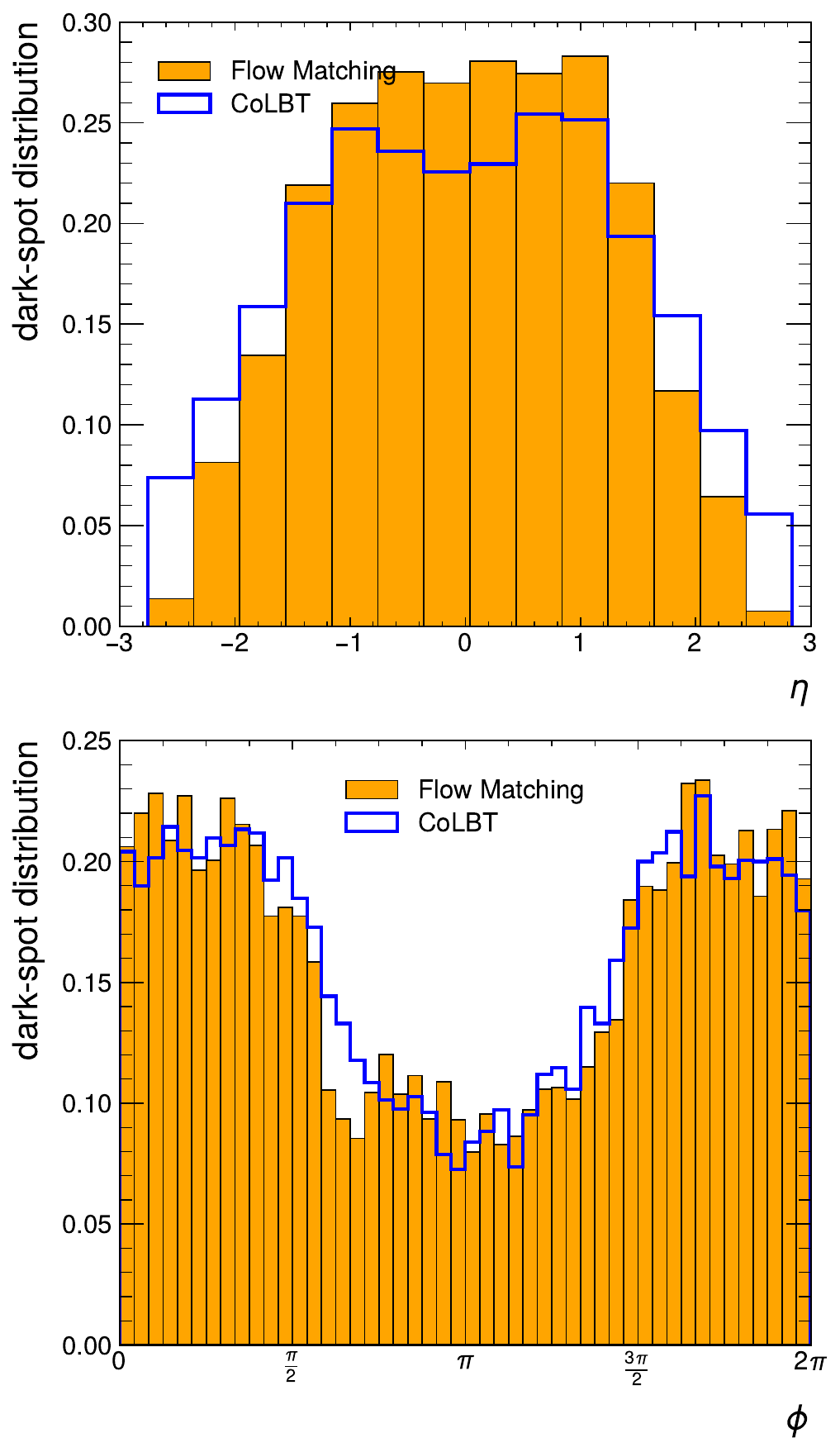}
    \caption{The location distributions of dark spots of diffusion wakes in rapidity $\eta$ (upper) and azimuthal angle $\phi$ (lower) from the Flow Matching model (shaded histograms) as compared to the CoLBT-hydro (open histograms) simulations.}
    \label{3Ddiffusionwake}
\end{figure}

Following the same analysis, we show in Fig.~\ref{3Ddiffusionwake} the location distributions of dark spots caused by the diffusion wakes in $\eta$ (upper panel) and $\phi$ (lower panel) from the Flow Matching model (shaded histogram) as compared to the CoLBT-hydro results (open histogram).  Naively, one would expect that the dark spots due to diffusion wake align with the hot spots or jet direction in rapidity and opposite to the hot spots in the azimuthal angle. However, radial flow and gradient of the medium can cause the diffusion wake to be deflected, leading to a subtle double-peak structure in both pseudo-rapidity and azimuth angle around $\eta=0$ and $\phi=0$ (the direction of the photon), as seen in both CoLBT-hydro and Flow Matching model results. Again the distributions from the Flow Matching model are slightly narrower than the training data due to the absence of the rare events with multiple sub-jets in the Flow Matching model.

\begin{figure}[h]
    \centering
    \includegraphics[width=0.8\linewidth]{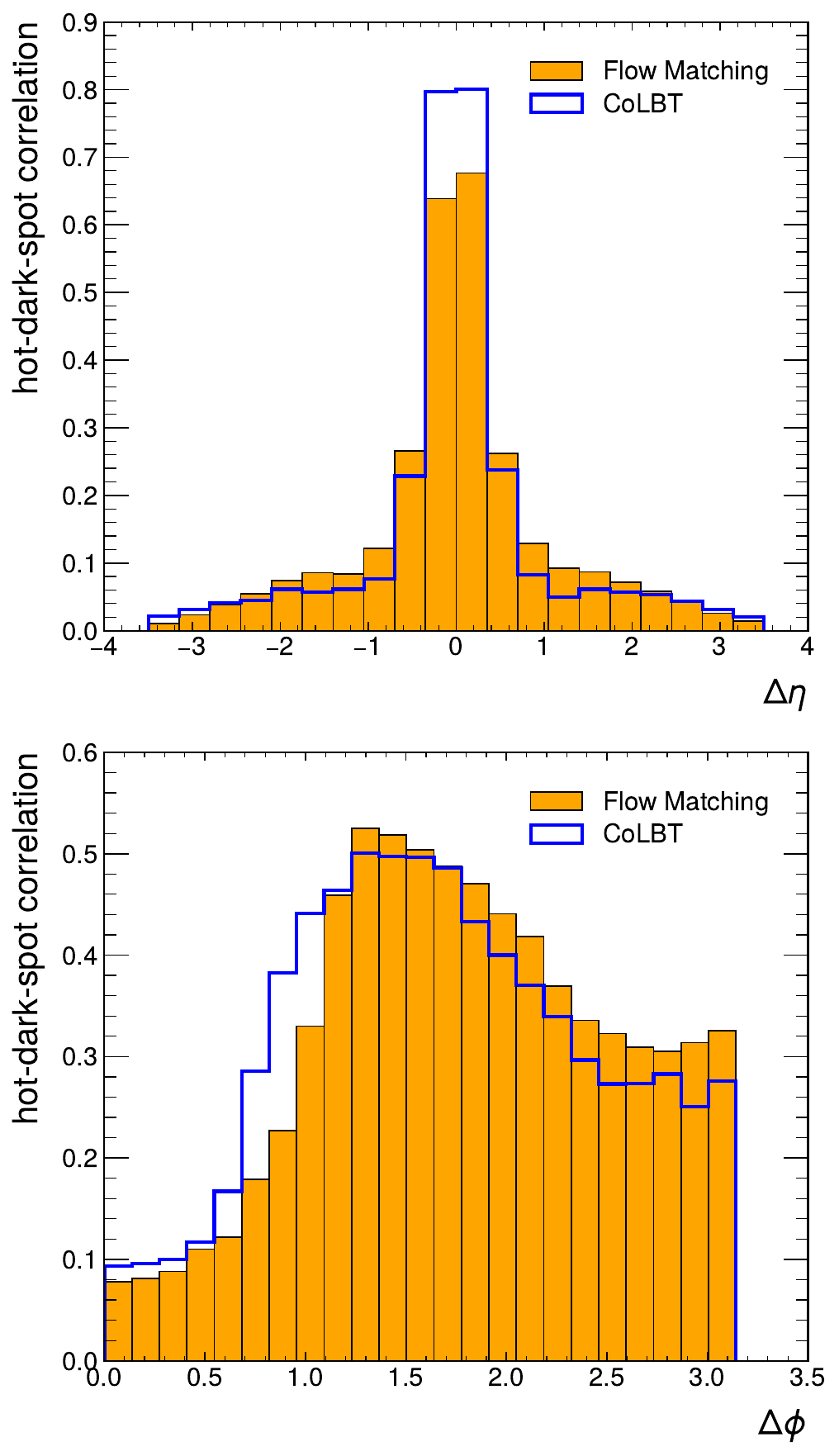}
    \caption{The correlations between hot (Front wake) and dark (Diffusion wake) spot in $\Delta\eta=\eta_{\rm hot}-\eta_{\rm dark}$ (upper) and $\Delta\phi=\phi_{\rm hot}-\phi_{\rm dark}$ (lower) from Flow Matching model (shaded histograms) as compared to the CoLBT-hydro (open histograms) simulations.}
    \label{3Ddeltaetaphi}
\end{figure}

To quantify the correlation between the locations of hot spots of the front wake and dark spots of the diffusion wake, we show in Fig.~\ref{3Ddeltaetaphi} the distributions of the distance between the locations of the hot and dark spots in pseudo-rapidity ($\Delta\eta$) (upper panel) and azimuthal angle ($\Delta\phi$) (lower panel) in each event. 

The distribution in $\Delta\eta$ peaks near $\Delta\eta=0$, as expected. However, the width is much narrower than the individual distribution of hot and dark spots, indicating very tight correlation between hot and dark spots in pseudo-rapidity in each event. The azimuthal angle difference $\Delta\phi$ distribution in the lower panel, in contrast, shows a strong double-peak structure. This implies that the minima of the particle depletion caused by the diffusion wake have a broad distribution in azimuthal angle and the peak is not exactly in the opposite direction of the hot spots (or jet direction).  The slight discrepancy between the Flow Matching and CoLBT-hydro model is again caused by the absence of events with multiple sub-jets in the Flow Matching model. 

\subsection{Hadron spectra from jet-induced hydro response}
\begin{figure}[h]
    \centering
    \includegraphics[width=0.7\linewidth]{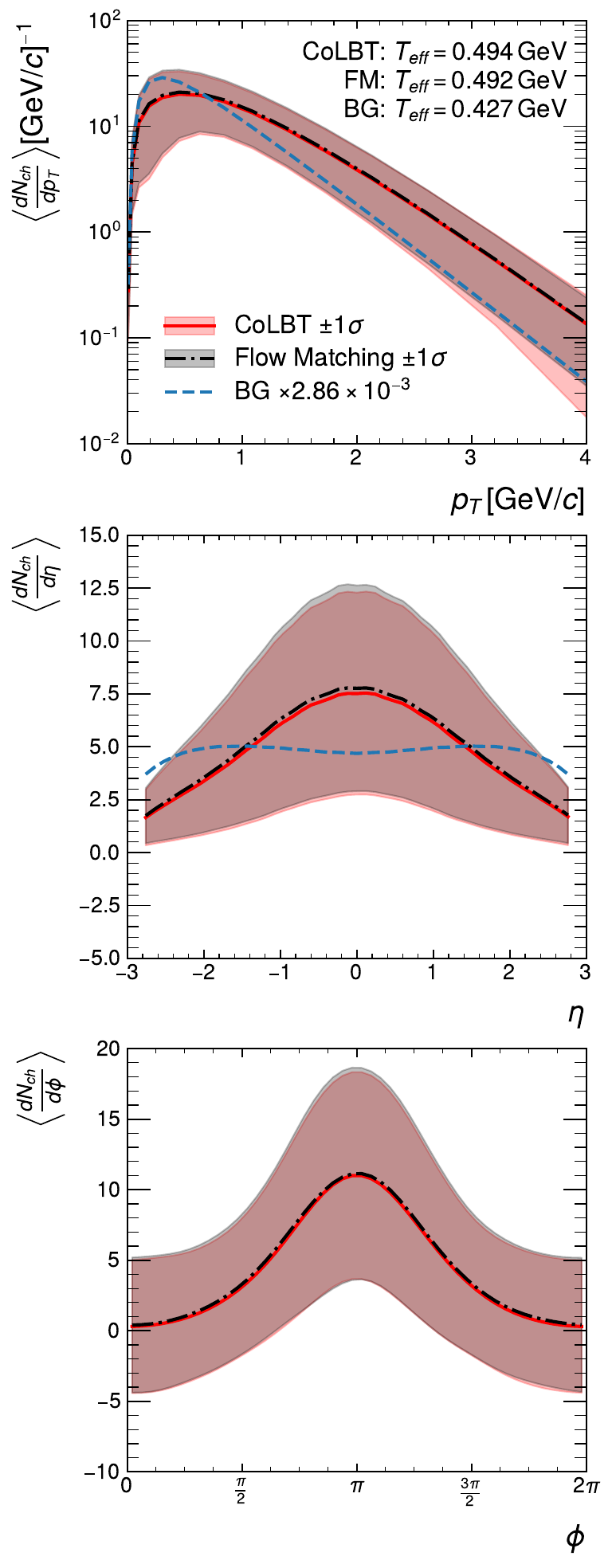}
    \caption{The event-averaged transverse momentum $p_T$ (upper), pseudo-rapidity $\eta$ (middle) and azimuthal angle $\phi$ (lower) distributions of the hydro response charged hadrons from CoLBT-hydro simulations (red solid) and Flow Matching model generations (black dot-dashed). Spectra from the background hydro without jets (blue dashed) are also plotted for comparison.  The effective temperatures $T_{\mathrm{eff}}$ are extracted by fitting the large $p_T\in[1,4]~\mathrm{GeV}/c$ region of the $p_T$ spectra with $dN/dp_T=A\,p_T\exp(-p_T/T_{\mathrm{eff}})$.}
    \label{3Dmean_line}
\end{figure}

To examine the inclusive hadron spectra in detail, we project the averaged differential 3D spectra $d^3N/dp_T d\eta d\phi$ onto the transverse momentum 
$p_T$, pseudo-rapidity $\eta$ and azimuthal angle $\phi$ as shown in Fig.~\ref{3Dmean_line}. The shaded bands represent statistical uncertainties at the 1-$\sigma$ level. The spectra from the Flow Matching model (dashed lines)
compare very well to the CoLBT simulations (solid lines) illustrating the effectiveness of the generative model to capture the essential features of the spectra from hydro response induced by $\gamma$-jets in central Pb+Pb collisions at $\sqrt{s_{\rm{NN}}}$ = 5.02 ~TeV. We also plot the spectra from the bulk background without $\gamma$-jets scaled by a factor $2.86\times 10^{-3}$ (blue dashed lines) for comparison. 

The transverse momentum ($p_T$) distribution of the hydro response resembles that of the
thermal background hadron spectra with hydrodynamic expansion in the upper panel.  However, the slope of the exponential hydro response spectra is harder than the background spectra due to jet energy loss deposited into the medium. To quantify the difference between the $p_T$ spectra of the hydro response and the background, we fit the large-$p_T$ region $p_T \in [1,4]~\mathrm{GeV}/c$ of the spectra with a thermal spectrum,
\begin{align}
    \frac{dN}{dp_T}=A\,p_T\exp\left(-\frac{p_T}{T_{\mathrm{eff}}}\right),
\end{align}
where $A$ is a normalization constant and $T_{\mathrm{eff}}$ is the effective temperature. The extracted $T_{\mathrm{eff}}$=0.494 and 0.492 GeV for the hydro response spectra from CoLBT-hydro and the Flow Matching model, respectively, are larger than $T_{\mathrm{eff}}$=0.427 GeV for the bulk background spectrum. Since the hydro response arises from the jet energy loss that is deposited into the medium, its averaged local temperature should be higher than the medium leading to the harder exponent of the final hadron $p_T$ spectra at freeze-out.

Note that we only consider the hydrodynamical response here whose exponential $p_T$ spectra at large $p_T$ are significantly smaller than the spectra of high-$p_T$ final-state hadrons from the hadronization of the hard partons of the medium-modified jet \cite{Chen:2017zte,Chen:2020tbl}.

The widths of the pseudo-rapidity ($\eta$) (middle panel) and azimuthal angle ($\phi$) (lower panel) distributions are determined both by the $\gamma$-jet correlation in $\eta$ and $\phi$ and hadron-jet correlation. Note again that we rotate the event plane such that $\gamma$ is always at $\phi=0$. The bulk background hadron spectrum is uniform in $\phi$ and the $\eta$ distribution is much wider than the hadrons from the medium response.

\begin{figure}[h]
    \centering
 \includegraphics[width=0.8\linewidth]{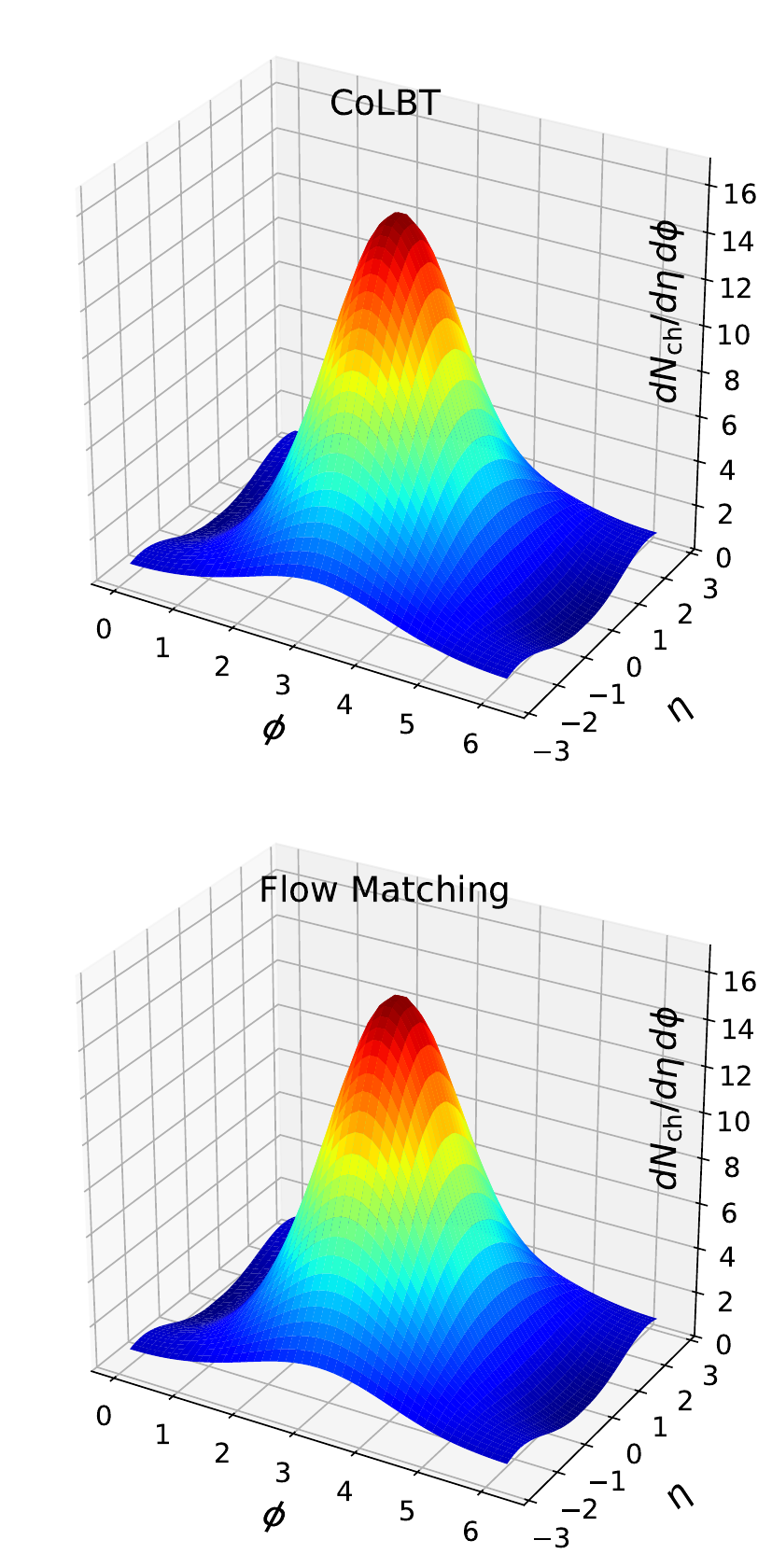}
    \caption{Event-averaged two-dimensional charged hadron spectrum $d^2N_{\rm ch}/d\eta d\phi$ of the hydro response induced by $\gamma$-jets from the CoLBT-hydro (upper) and Flow Matching model (lower) generations. The direct photon ($\gamma$) is set at $\phi_\gamma=0$.}
    \label{3Dmean_show}
\end{figure}

Since we project the 3D hadron spectra $d^3N/dp_T d\eta d\phi$ to one dimension in Fig.~\ref{3Dmean_line}, it is difficult to see the effect of the diffusion wake. As pointed out in earlier studies \cite{Yang:2022nei}, one can only see the effect of diffusion wake in 2D spectra $d^2N/d\eta d\phi$. Shown in Fig.~\ref{3Dmean_show} are these 2D distributions from CoLBT-hydro (upper) and the Flow Matching model (lower panel) after integration over $p_T$. The diffusion wake causes a depletion of hadrons in the opposite direction of the jet or along the photon direction. This leads to a double peak structure in the $\eta$ direction \cite{Yang:2022nei}. The multiple parton interaction (MPI) associated with jet production gives rise to a ridge along the $\phi$ direction and the diffusion wake associated with the jet-induced Mach cone produces a valley on top of the MPI ridge~\cite{Chen:2021gkj,Yang:2022nei}. The Flow Matching model has captured this structure due to the interplay of the front wake, diffusion wake and MPI ridge very well. 

\subsection{Diffusion wake and rapidity asymmetry}

The 2D spectra in $\eta$ and $\phi$ from jet-induced medium response in the last subsection are obtained by subtracting the background spectra (events without jets) from the spectra that contain contributions from both jet-induced medium response and the background. In experiments this amounts to a dedicated background subtraction which is very difficult and complicated \cite{ATLAS:2024prm,CMS:2025dua}. A new observable called rapidity asymmetry was proposed \cite{Yang:2025xni,Yang:2025lii} as a robust signal of the jet-induced diffusion wake that is also background-free. In $\gamma$-jet events in heavy-ion collisions, this rapidity asymmetry is defined as the difference between $\gamma$-associated hadron spectra,
\begin{align}
A(\eta_h)=\frac{dN}{d\eta_h}\Big|_{C_1}-\frac{dN}{d\eta_h}\Big|_{C_2},
\end{align}
of two event classes $C_1$ and $C_2$ with different jet-rapidity selections. Since the underlying hydrodynamic background is independent of jet's rapidity, contributions to the spectra from the background completely cancel. What is left is only the difference of the jet-induced medium response in these two classes of events. If we restrict the azimuthal angle to the photon direction, the above observable is essentially the difference between the diffusion wakes in the two classes of events with different rapidity of the diffusion wakes which follows that of the jet. The resultant rapidity asymmetry, therefore, provides a background-free signal of the jet-induced diffusion wake.

\begin{figure}
    \centering
    \includegraphics[width=0.8\linewidth]{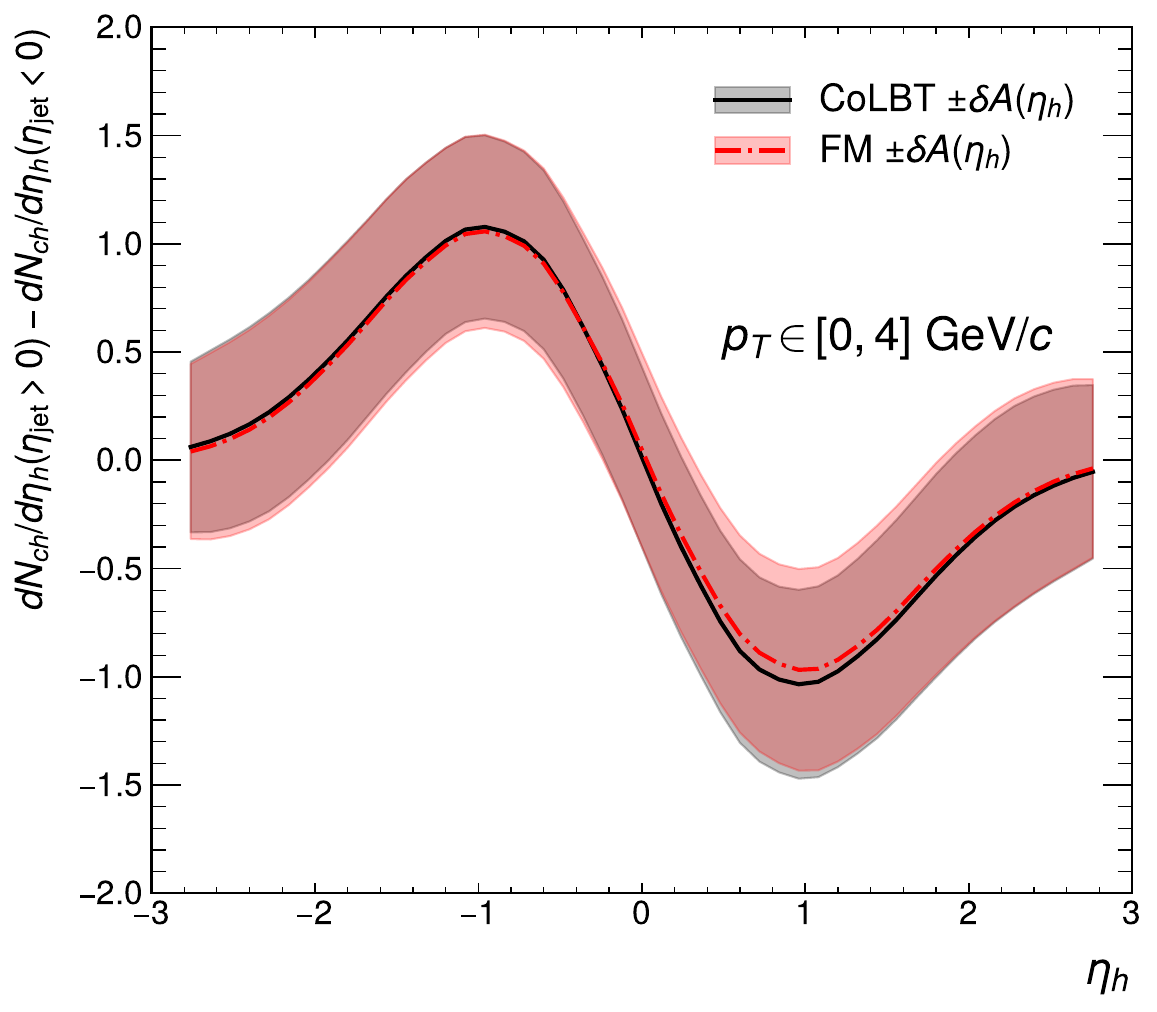}
    \caption{Rapidity asymmetry of the hydro-response charged hadrons induced by $\gamma$-jets as a function of $\eta_h$ for $p_T\in[0,4]$ GeV/$c$ and $ |\phi_h-\phi_\gamma|<\pi/2$ from the CoLBT-hydro (black solid) and the Flow Matching model generations (red dot-dashed). The shaded bands correspond to the statistical uncertainty $\pm \delta A(\eta_h)$.} 
    \label{Rap_asy2}
\end{figure}

Shown in Fig.~\ref{Rap_asy2} are the rapidity asymmetries between the event classes with $\eta_{\rm jet}>0$(class 1) and $\eta_{\rm jet}<0$(class 2) from both CoLBT-hydro (solid) and the Flow Matching model (dot-dashed). The azimuthal angle of the hadron is restricted to $|\phi_h-\phi_\gamma|<\pi/2$. The shaded bands are statistical errors.

As we have discussed earlier, both the front and the diffusion wake follow the jet in rapidity while the azimuthal angles of the diffusion wakes fall in the photon direction. The defined rapidity asymmetry is negative in the $\eta_{\rm h} >0$ region due to the
diffusion wake induced by jets with $\eta_{\rm jet} >0$, and positive in the $\eta_{\rm h} <0$ region due to the
diffusion wake induced by jets with $\eta_{\rm jet} < 0$. The asymmetry is antisymmetric in rapidity and
approaches zero at large rapidities when the effect of the diffusion wake disappears. The Flow
Matching model successfully reproduces this characteristic structure and is in good agreement with the CoLBT-hydro training data.

\subsection{Dependence on jet energy loss}

To further test the generative model's ability to capture the varying degrees of energy loss and its effect on the medium response, we use the $\gamma$-jet asymmetry,
\begin{equation}
     X^{\rm jet}_{\gamma}=\frac{p_T^{\rm jet}}{p_T^{\gamma}}
\end{equation}
to classify events, where $p_T^\gamma$ is the transverse momentum of the initial photon generated by PYTHIA 8 and $p_T^\text{jet}$ is the transverse momentum of the final-state jet reconstructed from the final-state hadrons using the anti-$k_T$ FastJet algorithm \cite{Cacciari:2011ma}. These hadrons are produced through the hadronization of the jet partons after they have traversed the quark-gluon plasma (QGP) in the CoLBT-hydro model. In principle, one should include hadrons from the medium response in the final-state hadronic jet reconstruction for a complete study. Since the energy carried by hadrons from the medium response inside the jet cone is small as compared to the total jet energy, we can neglect their contribution in this qualitative study of the Flow Matching model. As discussed in the section about the training dataset, a $p_T$ cut of 100 GeV is applied to the direct photon and a $p_T$ cut of 45 GeV is applied to the reconstructed jets. The larger this $\gamma$-jet asymmetry, $X^{\rm jet}_{\gamma}$, the higher the energy of the jet after traversing the QGP, and consequently, the smaller the energy loss. Conversely, a smaller value of this $\gamma$-jet asymmetry indicates a larger energy loss.

\begin{widetext}

\begin{figure}[h]
    \centering
 \includegraphics[width=0.8\linewidth]{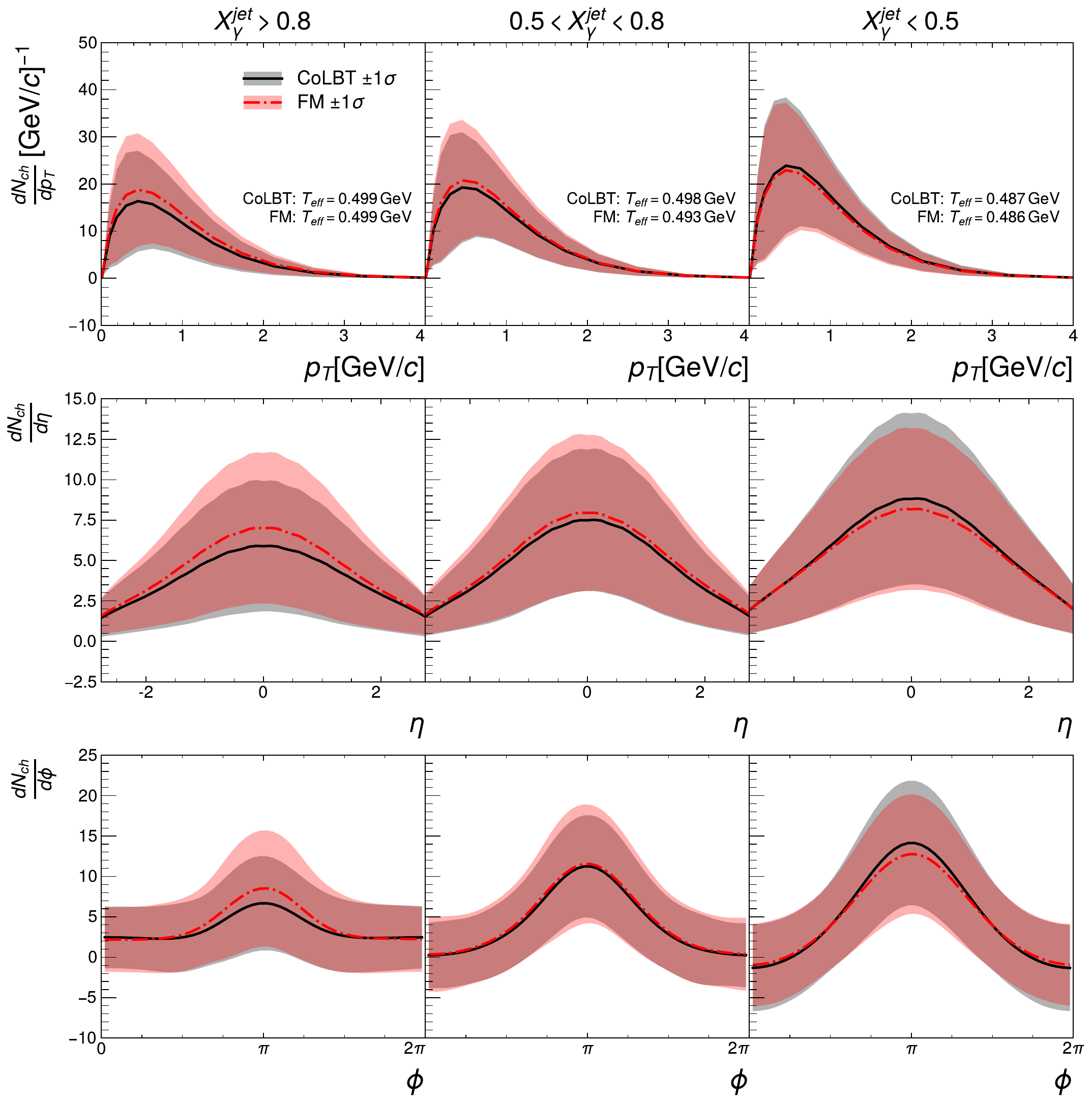}
 
    \caption{ The same as Fig.~\ref{3Dmean_line} except for different classes of events with different values of the $\gamma$-jet asymmetry $X_{\gamma}^{\mathrm{jet}}$}
    \label{fig:singlerange1}
\end{figure}

\end{widetext}

Shown in Fig.~\ref{fig:singlerange1} are hadron spectra from the medium response as a function of $p_T$ (upper panels), $\eta$ (middle panels) and $\phi$ (lower panels) for three different classes of events classified by the value of the $\gamma$-jet asymmetry: $X^{\text{jet}}_{\gamma} > 0.8$ (left column), $0.5 < X^{\text{jet}}_{\gamma} < 0.8$ (middle column), and $X^{\text{jet}}_{\gamma} < 0.5$ (right column). Solid black lines are results from the CoLBT-hydro simulations while the dot-dashed red lines are results generated by the Flow Matching model. The bands represent statistical uncertainties at the 1-$\sigma$ level.

One can easily notice the energy-loss dependence of the hadron multiplicity from the medium response. As the $\gamma$-jet asymmetry $X^{\text{jet}}_{\gamma}$ decreases,  the hadron multiplicity from the medium response increases as a result of the increased jet energy loss that is deposited into the medium. The effective temperature from fitting the exponential part of the $p_T$ spectra, however, decreases slightly when $X^{\text{jet}}_{\gamma}$ becomes smaller, possibly due to the long jet propagation length and the long evolution time of the medium response. The widths of the rapidity and azimuthal angle distributions are approximately independent of the $\gamma$-jet asymmetry $X^{\text{jet}}_{\gamma}$. The Flow Matching model reproduces the CoLBT-hydro results well within the statistical errors, although the generative model overestimates the overall hadron multiplicity for large $X^{\text{jet}}_{\gamma}>0.8$ (small jet energy loss) and underestimates the hadron multiplicity for small $X^{\text{jet}}_{\gamma}<0.5$ (large jet energy loss). This leaves room for future improvement of the generative model, for example, by introducing additional categorization of events by the jet energy loss in the training dataset.

\section{Conclusions}
In this study, we have developed a generative neural network model to accelerate the Monte Carlo simulations of jet-induced hydro response in relativistic heavy-ion collisions. On an NVIDIA RTX 4090 GPU, the generative model achieves a speedup of approximately six orders of magnitude in generating the final-state particle spectra from the hydro response as compared to the conventional CoLBT-hydro model. Requiring only the initial photon ($\gamma$)- jet configuration as input, the model can directly generate the complete three-dimensional final-state particle spectra $d^3N/dp_Td\eta d\phi$ from the hydro response induced by a jet.

We have systematically validated the model's performance through quantitative comparisons, demonstrating that the generated particle spectra are in good agreement with the training CoLBT-hydro model. Due to the fluctuations of the initial hydro conditions and fluctuations intrinsic to the multiple scattering and induced branching in the QGP medium, the generative model cannot reproduce the CoLBT-hydro model on an event-by-event basis. It can, however, reproduce the event-averaged hadron spectra, both the internal structure and the correlations, for example the back-to-back spatial correlation between the front and the diffusion wake, the jet-energy-loss dependence of the hadron multiplicity from the jet-induced hydro response. The generative model can also faithfully reproduce the valley on the MPI ridge in the 2D hadron spectra $d^2N/d\eta d\phi$ and the rapidity asymmetry caused by the diffusion wake.

This study can be considered as a proof of principle for AI generation of high-energy heavy-ion collisions and jet propagation with the Flow Matching model. Though there is room for future improvement, for example in reproducing events with multiple sub-jets and the precise energy-loss dependence of the hadron multiplicity from the hydro response, our study demonstrates an efficient generative model for simulations of jet-induced hydro response directly from jet initial configurations. The framework possesses excellent scalability. It can be applied to other collision systems, such as dijet or Z-jet events, by simply retraining the model on the appropriate datasets without modifying the underlying network architecture. We only focused on the $0-10\%$ central Pb+Pb collisions at $\sqrt{s_{\rm NN}}=5.02$ TeV in this study. By tagging the initial hydro conditions such as the initial entropy  density distribution, one can incorporate the centrality and collision-energy dependence. The model can also be extended to include the event-by-event hydrodynamic evolution of the bulk medium as well as the transport of hard jet shower partons. This opens up broad prospects for applying advanced machine learning techniques in event-by-event simulations of jet transport and dynamical evolution of the dense matter in high-energy heavy-ion collisions.

\section{Acknowledgements}
This work is supported in part by the Postdoctoral Fellowship Program and China Postdoctoral Science Foundation under Grant No. BX20240134 (ZY) and China Postdoctoral Science Foundation under Grant No. 2024M751059 (ZY), by NSFC under Grant No. 12535010, No. 12075098, No. 12435009. Computations in this work were carried out at the Nuclear Science Computing Center at Central China Normal University (NSC3).

\bibliography{newref}  

\end{document}